\documentclass[pre,aps,superscriptaddress]{revtex4}
\usepackage{graphicx}
\usepackage{amsmath,amsfonts}

\newcommand{\beq}{\begin{equation}}
\newcommand{\eeq}{\end{equation}}
\newcommand{\bea}{\begin{eqnarray}}
\newcommand{\eea}{\end{eqnarray}}
\newcommand{\re}{\text{Re}}
\newcommand{\im}{\text{Im}}
\newcommand{\ie}{{\it i.e.} }
\newcommand{\eg}{{\it e.g.} }
\newcommand{\T}{{\tau}}
\newcommand{\su}{\sigma^{\Th}}
\newcommand{\Iint}{\int_{-\infty}^\infty}

\let\a=\alpha \let\b=\beta  \let\g=\gamma  \let\d=\delta 
\let\z=\zeta     \let\th=\theta \let\k=\kappa \let\l=\lambda
\let\m=\mu    \let\n=\nu         \let\p=\pi    \let\r=\rho
\let\s=\sigma \let\t=\tau   \let\f=\varphi 
  \let\y=\upsilon \let\o=\omega
 \let\D=\Delta  \let\Th=\Theta 
    \let\Si=\Sigma     
\let\O=\Omega 

\let\io=\infty

\let\la=\langle
\let\ra=\rangle
\let\dpr=\partial

\begin{document}

\title{Fluctuation theorem for non-equilibrium
relaxational systems \\ driven by external forces
}

\author{
Francesco~Zamponi\footnote{francesco.zamponi@phys.uniroma1.it}
}
\affiliation{
Dipartimento di Fisica, Universit\`a di Roma {\em La Sapienza},
P.le A. Moro 2, 00185 Roma, Italy
}
\affiliation{
SOFT--INFM--CNR, Universit\`a di Roma {\em La Sapienza},
P.le A. Moro 2, 00185 Roma, Italy
}
\author{
Federico Bonetto\footnote{bonetto@math.gatech.edu}
}
\affiliation{
School of Mathematics, Georgia Institute of Technology,
 Atlanta, GA 30332 
}
\author{
Leticia~F.~Cugliandolo\footnote{leticia@lpt.ens.fr}
}
\affiliation{
Laboratoire de Physique Th\'eorique et Hautes \'Energies
  Jussieu,
 4 Place Jussieu,  75252 Paris Cedex 05, France
}
\affiliation{
Laboratoire de Physique Th\'eorique, 
\'Ecole Normale Sup\'erieure, 24 Rue Lhomond,
75231 Paris Cedex 05, France
}
\author{
Jorge~Kurchan\footnote{jorge@pmmh.espci.fr}
}
\affiliation{
P.M.M.H. \'Ecole Sup\'erieure de Physique et 
Chimie Industrielles, 10 Rue Vauquelin, 75231 Paris
 Cedex 05, France
}

\begin{abstract}
We discuss an extension of the fluctuation theorem to stochastic
models that, in the limit of zero external drive, are not able to
equilibrate with their environment, extending results presented by
Sellitto (cond-mat/9809186).  We show that if the entropy production
rate is suitably defined, its probability distribution function
verifies the Fluctuation Relation with the ambient temperature
replaced by a (frequency-dependent) effective temperature. We derive
modified Green-Kubo relations.  We illustrate these results with the
simple example of an oscillator coupled to a nonequilibrium bath
driven by an external force. We discuss the relevance of our results
for driven glasses and the diffusion of Brownian particles in out of
equilibrium media and propose a concrete experimental strategy to
measure the low frequency value of the effective temperature using the
fluctuations of the work done by an ac conservative field. We compare
our results to related ones that appeared in the literature recently.
\end{abstract}

\maketitle


\section{Introduction}
\label{sect:intro}
Relatively few generic results for non-equilibrium systems exist.  Recently,
two such results that apply to seemingly very different physical situations
have been proposed and extensively studied.  One is the {\it fluctuation
theorem} that characterizes the fluctuations of the entropy production over
long time-intervals in
certain driven steady states~\cite{ECM93,GC95a}.  Another one is the extension
of the fluctuation--dissipation theorem that relates induced and spontaneous
fluctuations in equilibrium to the non-equilibrium slow relaxation of glassy
systems~\cite{Cuku93,Cuku}. 
While the former result has been proven for
reversible hyperbolic dynamical systems~\cite{GC95a,GC95b} and 
for the driven stochastic dynamic evolution of an open system coupled to an
external environment~\cite{Ku98,LS99}, the latter has only been obtained in a
number of solvable mean-field models and numerically in some more realistic
glassy systems (see~\cite{Felix-review} for a review).  
The modification of the fluctuation--dissipation theorem can
be rationalized in terms of the generation of an {\it effective
temperature}~\cite{Cukupe,Cukuja}.  
The expected thermodynamic properties of the
effective temperature have been demonstrated in a number of
cases~\cite{proviso}.

One may naturally wonder whether these two quite generic results
may be included in a common, more general statement.
 The scope of this article is 
to discuss this possibility in general, illustrating it with
a  very simple  example
in which one can very easily reach the `driven limit' and the 
`non-equilibrium relaxational' case. This project was pioneered by 
Sellitto~\cite{Se98} who asked the same question some years ago
and tried to give it an answer using a stochastic lattice gas with 
reversible kinetic constraints in diffusive contact with two particle 
reservoirs at different chemical potentials. Other developments 
in similar directions have been proposed and analyzed by 
several authors~\cite{SCM04,CR04,Sasa}. We shall discuss them in 
Sect.~\ref{sec:conclusions}.

Before entering into the details 
of our calculations let us start by reviewing the precise statement 
of the usual fluctuation theorem and the extension of the
fluctuation--dissipation theorem as well as the definition of the 
effective temperature.

\subsection{The fluctuation theorem} 
\label{sect:intro-ft} 
 
The {\it fluctuation theorem} concerns the fluctuations of the entropy
production rate~\cite{comment1},
that we call $\sigma(x)$, in the stationary
non-equilibrium state of a dynamical system defined by a state variable $x \in
M$, where $M$ is the phase space of the system.
 
In non-equilibrium  
stationary states the function $\sigma(x)$ has a positive average,  
$\sigma_+ = \int_M dx \m_+(x) \s(x)>0$, where $\m_+(x)$ is the stationary state
distribution. This allows one to define  
\beq 
\label{pdef} 
p(x) \equiv\frac{1}{\T \sigma_+}{\cal{S}}_\t
= \frac{1}{\T \sigma_+}\int_{-\T/2}^{\T/2} dt \ \sigma(S_t x) \ , 
\eeq  
where 
$S_t x$, with time $t \in [-\T/2,\T/2]$, is a segment of the system's
trajectory starting at the point $x$ at $t=-\T/2$.  
The {\it large deviation function} of $p(x)$ is 
\beq  
\zeta(p) = \lim_{\T 
\rightarrow \infty} \T^{-1} \ln \pi_\T(p) \ ,
\label{eq:z-def}  
\eeq  
where $\pi_\T(p)$  is the distribution of $p$ in the stationary non-equilibrium state. 
The fluctuation theorem is the following statement about the  
large deviation function of $p(x)$:
\beq  
\zeta(p)-\zeta(-p)=p\sigma_+ \ .   
\label{eq:rel-ft}
\eeq 
 
The relation (\ref{eq:rel-ft}) was first discovered in a numerical 
simulation~\cite{ECM93}, and subsequently stated as a theorem for 
reversible hyperbolic dynamical systems (the {\it Gallavotti-Cohen 
Fluctuation Theorem})~\cite{GC95a,GC95b}. The proof was extended to 
Langevin systems coupled to a white bath in \cite{Ku98} and to generic 
Markov processes in \cite{LS99}. Equation~(\ref{eq:rel-ft}) was 
successfully tested in a wide number of numerical simulations, see \eg
\cite{BGG97,BCL98,GRS04,ZRA04a}, and, more recently, it was also
analyzed in experiments on 
granular materials and turbulent flows \cite{CL98,GGK01,CGHLP03,FM04}. 
At present, it is believed to yield a very general 
characterization of the fluctuations of the entropy production 
in a variety of out of equilibrium stationary states. 
 
In all the cases cited above 
an external drive maintains the systems 
in a stationary non-equilibrium regime. In the absence of the drive  
the systems so far analyzed easily equilibrate.  
In the stochastic case, the systems are in contact with an  
equilibrated environment at a well-defined temperature. 
 
\subsection{The effective temperature} 
\label{sect:intro-fdt-teff} 
 
The analytic solution to the relaxation of mean-field glassy models 
following a quench into their glassy phase  demonstrated that their 
relaxation occurs out of equilibrium~\cite{Cuku93,Cuku}. 
The reason why these models  
do not reach equilibrium when relaxing from a random initial  
condition is that their equilibration time diverges with  
their size. Thus, when the thermodynamic limit  
is taken at the outset of the calculation, all times considered  
are finite with respect to the size of the system. These 
systems approach a slow nonequilibrium regime in which one observes 
a breakdown of stationarity and, more importantly for the  
subject of this paper, a violation of the  
fluctuation-dissipation theorem that relates spontaneous and  
induced fluctuations in thermal equilibrium. 
 
The relation between spontaneous and induced fluctuations found in 
mean-field glassy models is, however, surprisingly simple.   
Let us define the linear response of a generic  observable $O$ 
measured at time $t$ to an infinitesimal perturbation constantly 
applied since a previous `waiting-time' $t_w$, and the     
correlation between the same (unperturbed) observable 
measured at $t_w$ and $t$,  
\begin{eqnarray} 
\chi(t,t_w) &\equiv&  
\int_{t_w}^t dt' \; R(t,t') = 
\int_{t_w}^t dt' \;  
\left. \frac{\delta \langle O(t) \rangle}{\delta h(t')} \right|_{h=0} 
\label{eq:chi-def}
\; , 
\\ 
C(t,t_w) &\equiv& \langle O(t) O(t_w)\rangle 
\; , 
\label{eq:C-def}
\end{eqnarray} 
where, for simplicity, we assumed that the observable $O$ has a vanishing 
average, $\langle O(t)\rangle =0$ for all $t$. 
In the cases we shall be interested in the relation between  
these two quantities takes the form 
\begin{equation} 
\lim_{t_w \gg t_0} \chi(t,t_w) = f[C(t,t_w)] 
\end{equation} 
in the long waiting-time limit after the initial time $t_0$.
This equation means that the waiting-time and total  
time dependence in $\chi$ enters only  
through the value of the associated correlation between these times. 
This is trivially true in equilibrium since  
the {\it fluctuation-dissipation theorem} states 
\begin{equation} 
\label{fCFDT}
f(C) = \frac{1}{T} \; (1-C) 
\; , 
\end{equation} 
for all times $t\geq t_w\geq t_{eq}$,  
where $T$ is the temperature of the thermal bath (and the one of the 
system as well) and  
we set the Boltzmann constant to one, $k_B=1$. 
Out of equilibrium $f(C)$ can take a different form. In a large  
variety of models 
with {\it slow dynamics}  
that describe aspects of glasses $f(C)$ is a broken line, 
\begin{equation} \label{twoT}
f(C) = \frac{1}{T} \; (1-C) \; \theta(C-q_{ea}) +  
\left[\frac{1}{T} \; (1-q_{ea}) + \frac{1}{T_{eff}} (q_{ea}-C) \right] 
\; \theta(q_{ea}-C) 
\; ,
\end{equation} 
see \cite{Felix-review} for a review.  This broken line has two
slopes, $-1/T$ for $C> q_{ea}$ ({\it i.e.}, small $t-t_w$), and
$-1/T_{eff}$ for $C< q_{ea}$ ({\it i.e.}, large $t-t_w$).  The
breaking-point $q_{ea}$ has an interpretation in terms of intra-cage
and out-of-the cage motion in the relaxation of structural glasses,
inter-domain and domain wall motion in coarsening systems, etc.  Since
$T_{eff}$ is found to be larger than $T$ the second term violates the
fluctuation-dissipation theorem. We have used the suggestive name {\it
effective temperature}, $T_{eff}$, to parametrize the second slope.
The justification is that for mean-field glassy models --- and within
all resummation schemes applied to realistic ones as well ---
$T_{eff}$ does indeed behave as a temperature~\cite{Cukupe}.  A
similar relation was later found numerically in the slow relaxational
dynamics of a number of more realistic glassy systems such as
Lennard-Jones mixtures~\cite{Lennard-FDT}. More complicated forms,
with a sequence of segments with different slopes appear in mean-field
glassy models of the Sherrington-Kirkpatrick type.

The general definition of the effective temperature is 
\begin{equation}\label{eq:Tefftime-def}
-\frac{1}{T_{eff}(C)} = \frac{d \chi}{d C} \; .
\end{equation}
In order to be consistent with the thermodynamic properties one 
needs to find a single value of $T_{eff}$ in each time
regime as defined by the correlation scales of Ref.~\cite{Cuku}.

So far we considered systems relaxing out of equilibrium {\it in
the absence of external drive}.  The extension of the definition of the
effective temperature to the {\it driven} dynamics of glasses and
super-cooled liquids was again motivated by the analytic solution of
mean-field models under non-potential
forces~\cite{Cukulepe,Cukulepe1,BBK00}. The existence of non-trivial effective
temperatures, {\it i.e.} of functions $f(C)$ that differ from
Eq.~(\ref{fCFDT}), was also observed in numerical simulations of
sheared Lennard-Jones mixtures~\cite{BB00,BB02} and
in a number of other driven low-dimensional models~\cite{lowd-driven}.

In the case of relaxing glasses the dynamics occurs out of equilibrium
because below some temperature the equilibration time
falls beyond all experimentally accessible time-scales. 
These macroscopic systems then evolve out of
equilibrium even if they are in contact with a thermal reservoir,
itself in equilibrium at a given temperature $T$.
Under the effect of stirring forces, 
 supercooled liquids and glasses  are typically driven
into a nonequilibrium stationary regime, even for relatively modest 
flow.

\subsection{A connection between the two?} 
\label{sect:intro-conn} 

The fluctuation theorem and the fluctuation--dissipation theorem are
related: indeed, it was proven that, for systems which are able to
equilibrate in the small entropy production limit ($\sigma_+
\rightarrow 0$), the fluctuation theorem implies the Green-Kubo
relations for transport coefficients, that are a particular instance
of the fluctuation--dissipation theorem~\cite{Ga96a,Ga96b,Ku98}.

It is now natural to wonder what is the fate of the fluctuation
theorem if the system on which the driving force is applied cannot
equilibrate with its environment and evolves out of equilibrium even
in the absence of the external drive.  Our aim is to
investigate whether the fluctuation theorem is modified and, more
precisely, whether the effective temperature~\cite{Cukupe} enters its
modified version. In particular, this question will arise 
if the limit of large sampling time, $\tau$ in 
Eqs.~(\ref{pdef}) and (\ref{eq:z-def}), is taken after the limit of 
large system size.  The order of the limits is important
because a finite size undriven 
system will always equilibrate with the thermal bath
in a large enough time $\t$. As the fluctuation theorem concerns the
fluctuations of $\s$ for $\t\to\io$, if one wants to observe 
{\it intrinsic nonequilibrium}
effects, the latter limit has to be taken {\it after}
the thermodynamic limit.
Some conjectures about this issue have recently
appeared in the literature~\cite{CR04,SCM04,Se98,Sasa} and we discuss
them in Sect.~\ref{sec:conclusions}.
 
Our idea is to study the relaxational and driven dynamics of  
simple systems such that the effective temperature is not trivially 
equal to the ambient temperature.  For a system coupled to a single 
thermal bath, this happens whenever: 
 
{\it (i)} the thermal bath has temperature $T$, but the system is not
able to equilibrate with the bath. This is realized by the glassy
cases discussed above, provided the sampling time is smaller than the
equilibration time; and/or
 
{\it (ii)} the system is very simple (not glassy) but it is set in contact  
with a bath that is not in equilibrium.  One can think of 
two ways of realizing this. One is with a single bath 
represented by a thermal noise and a memory friction kernel that do
not verify the fluctuation-dissipation relation \cite{Cu02}.  This 
situation arises if one considers the diffusion of a Brownian 
particle in a complex medium (\eg a glass, or granular matter) 
\cite{PM04,Po04,AG04}.  In this case the medium, which acts as a 
thermal bath with respect to the Brownian particle, is itself out of 
equilibrium.  Another example is the one of a  system coupled to a number 
of equilibrated thermal baths with different time-scales and at 
different temperatures~\cite{Cukuja}. 
 
\subsection{Effective equations for the dynamics of mean-field glasses}

The cases {\it (i)} and {\it (ii)} mentioned in the previous section
are closely related as, at least at the mean-field level, the problem
of glassy dynamics can be mapped onto the problem of a single
``effective'' degree of freedom moving in an out of equilibrium
self-consistent environment~\cite{Cu02,Cukuja}. Situations {\it (i)}
and {\it (ii)} are then described by the same kind of equation,
namely, a Langevin equation for a single degree of freedom coupled to
a non-equilibrium bath.

Indeed, in the study of mean-field models for glassy
dynamics~\cite{Cu02,Cukuja} and when using resummation techniques
within a perturbative approach to microscopic glassy models with no
disorder, it is possible to reduce the $N$-dimensional equations of
motion to a single equation for an `effective' variable by means of a
saddle-point evaluation of the dynamic generating functional.
These equations are
 valid in the large size $N\to\io$  limit;
 the discussion that follows concerns the fluctuation 
relation involving  ${\cal{S}}_\t$ for $\t \rightarrow \infty$ taken 
{\em after} $N\to\io$: for these finite timescales $\t$ a fluctuation
relation involving effective temperatures may and will arise, while
for the extreme times $\t \rightarrow \infty$ before  $N\to\io$ the usual
fluctuation theorem (involving only the bath temperature $T$) holds.

  The
effective equation of motion of a single spin at finite time reads 
\beq
\label{motionglassy} 
\g \dot \f_t = -\mu(t) \ \f_t + \int_{-\io}^t dt' \ \Si(t,t') \
\f_{t'} + \r_t + \a h[\f_t] 
\ , 
\eeq 
where $\r_t$ is a Gaussian noise
such that $\langle \r_t \r_{t'} \rangle = D(t,t')$.  The
``self-energy'' $\Si(t,t')$ and the ``vertex'' $D(t,t')$ depend on the
interactions in the system. They cannot be calculated exactly for
generic interactions but they can be approximated within different
resummation schemes (mode-coupling, self-consistent screening, etc.)
or calculated explicitly for disordered mean-field models.  The last
term (if present) represents an external drive.
 
In the absence of the driving force Eq.~(\ref{motionglassy}) shows a
{\it dynamical transition} at a temperature $T_d$.  Above $T_d$, in
the limit $t,t' \rightarrow \io$ the functions $\Si(t,t')$ and
$D(t,t')$ become time-translation invariant and are related by the
fluctuation-dissipation relation with $\Sigma$ playing the role of a
response and $D$ being a correlation. Below $T_d$ the system is no
more able to equilibrate with the thermal bath and {\it ages}
indefinitely, \ie the functions $\Si(t,t')$ and $D(t,t')$ depend on
$t$ and $t'$ separately even in the infinite time limit, and the
relaxation time $\t_\a$ grows indefinitely.  At low temperatures the
fluctuation-dissipation theorem does not necessarily hold, and the
relation between $\Sigma$ and $D$ can be used to measure the effective
temperature for the particular problem at hand.
 
In the case of a driven mean field
system~\cite{Cukulepe,Cukulepe1,BBK00}, the external force is also
present in Eq.~(\ref{motionglassy}) and after a transient the system
becomes stationary {\it for any temperature}, \ie $\mu(t) \equiv \mu$,
$\Si(t,t')=\Si(t-t')$, and $D(t,t')=D(t-t')$. The functions $D$ and
$\Si$ depend on the strength $\a$ of the driving force. Below $T_d$
and for small $\a$ they are again related by a generalization of the
fluctuation-dissipation theorem similar to the one obtained for
$\a=0$.

In this paper we focus on Langevin equations similar to
Eq.~(\ref{motionglassy}). Their main characteristic is that the
thermal bath, represented by the functions $\Si(t,t')$ and $D(t,t')$,
is not in equilibrium.  As we are interested in the driven case, we
restrict to considering stationary functions $\Si(t-t')$ and
$D(t-t')$.

Let us  remark that Eq.~(\ref{motionglassy}) is expected to
describe the dynamics of a single Brownian particle immersed in a
non--equilibrium environment, or the dynamics of an effective degree
of freedom representing a many--particle mean--field glassy system.
If one wishes to describe in full detail the relaxation of real
glasses in finite dimensions, more complicated effects have to be
taken into account.  For example, the dynamics is expected to be
heterogeneous yielding a {\it local} effective temperature which may
depend on space inside the sample, see e.g.~\cite{Cu04} for a detailed
discussion.  The extension of the results that we shall present in
this paper to glassy systems in finite dimension might require
additional work.
 
\subsection{Summary and structure of the paper} 
 
The aim of this work is to discuss the validity of the fluctuation
theorem for the Langevin equation~(\ref{motionglassy}) in presence of
a non-trivial environment represented by the functions $\Si(t)$ and
$D(t)$.  This equation describes a situation in which the relaxing
system does not equilibrate with its environment (even in the absence
of driving forces) and a non-trivial effective temperature defined
from the modification of the fluctuation-dissipation theorem exists.
Our first aim is to identify the entropy production rate
and to show that the effective temperature of the environment 
should replace the ambient temperature of a conventional 
bath. 

As particular cases we investigate analytically the dynamics
in a harmonic well and numerically a case in which the equations of
motion are nonlinear. The former problem is relevant for experiments
on confined Brownian particles in complex media \cite{AG04,Gr03}.
In both cases we show that we show that the entropy production rate 
that we introduced verifies the fluctuation relation in general.  

The paper is organized as follows.  In section~\ref{sec:LS} we
identify the correct definition of entropy production rate using a
rather general procedure proposed by Lebowitz and Spohn.  As an
illustration, in section~\ref{sec:II} we analytically compute the
large deviation function in the harmonic case for a white equilibrium
bath, already discussed in Ref.~\cite{Ku98}, and for a complex bath.
We show that the large deviation function is a convex function of $p$
and satisfies the fluctuation relation.  In section~\ref{sec:V} we
compare the analytic results obtained for the linear problem with
numerical data. We also numerically investigate a nonlinear Langevin
equation, in which the entropy production is not only given by the
work of the external forces, but should contain an `internal' term as
well. Surprisingly, this term turns out to be negligible, a result we
attribute to decorrelation between the work of internal and external
forces.  In section~\ref{sec:VI} we make contact with glassy problems
and discuss some connections with recent numerical
simulations. Section~\ref{sec:Green-Kubo} is devoted to the analysis
of the link between the modified fluctuation theorem and modified
Green-Kubo relations. In Section~\ref{sec:XI} we show explicitely that
the distribution of the work done by a slow periodic drive satisfies
the fluctuation relation with the low-frequency effective temperature,
and argue that this procedure is the one that could be ``easily''
implemented experimentally as a means to measure the (low frequency)
effective tempeature.  In the conclusions we briefly discuss some
experiments that could test our predictions, and compare them to
previous studies of the same problem.

 
\section{Entropy production}
\label{sec:LS}

In this Section we introduce the set of dynamic
models that we discuss in detail, namely, stochastic processes of  
generic Langevin type with additive noise. Next, we define the large deviation 
function that we shall use to test the validity of the fluctuation 
relation and, finally, we derive a general expression for 
the entropy production rate.

\subsection{The model}

In this article we focus on different aspects of the random motion of
a particle in a confining potential, in contact with a thermal
environment, and under the effect of a driving external force.  The
Langevin equation describing the motion of such a particle in a $d$
dimensional space reads
\begin{equation} 
m \ddot r_\a(t) + \Iint dt' \; g_{\a\b}(t-t') \dot r_\b(t') =  
-\frac{\partial V}{\partial r_\a}[\vec r(t)] +  
\rho_\a(t) 
+ h_\a(t)
\; , 
\;\;\;\;\;\;\;\;\; 
\a=1,\dots, d 
\; . 
\label{eq:linear-t-harm} 
\end{equation} 
${\vec r}=(r_1,\dots,r_d)$ is the position of the particle.  We pay
special attention to the case $d=2$ and call $(x,y)$ the two
components of the position vector.  $m$ is the mass of the particle
and $V({\vec r})$ is a potential energy. As an example we shall work
out in detail the simple harmonic case, $V({\vec r})=\frac{k}{2}
\sum_\a r_\a^2$ with $k$ the spring constant of the quadratic
potential.  ${\vec \rho}(t)$ is a Gaussian thermal noise with zero
average and generic stationary correlation
\begin{equation} 
\langle \, \rho_\a(t) \rho_\b(t') \, \rangle = 
\delta_{\a\b} \; \nu(t-t') 
\;\;\;\;\;\;\;\;\; 
\a,\b=1,\dots, d 
\; , 
\end{equation} 
with $\nu(t-t')$ a symmetric function of $t-t'$.  The memory kernel
$g_{\a\b}(t-t')$ extends the notion of friction to a more generic
case. We assume a simple spatial structure, $g_{\a\b}(t-t') =
\delta_{\a\b} \; g(t-t')$. In order to ensure causality we take
$g(t-t')$ to be proportional to $\theta(t-t')$. We define
\begin{equation}
g(t) = \theta(t)  f(t) \qquad \mbox{with} 
\qquad
f(t) \equiv g(t) + g(-t) 
\; .
\label{f-def}
\end{equation}

The initial time $t_0$ has been taken to $-\infty$. 
${\vec h}(t)$ is a time-dependent field that we either use to compute 
the linear response or represents the external forcing.

It will be useful to introduce Fourier  transforms.
We use the conventions 
\begin{equation} 
\rho(t) = \Iint 
\frac{d\omega}{2\pi} \; e^{-i\omega t} \, \rho(\omega) 
\; ,  
\;\;\;\;\;\;\;\;\;\;\;\;\;\; 
\rho(\omega) = \Iint dt \; e^{i\omega t} \, \rho(t) 
\; . 
\end{equation} 
 
The fluctuation-dissipation theorem states that for systems evolving 
in thermal equilibrium with their equilibrated environment the linear 
response is related to the correlation function of the same observable 
as 
\begin{equation} 
R(t) = - \frac{\theta(t)}{ T} \frac{dC(t)}{dt} \; , 
\;\;\;\;\;\;\;\;\;\;\;\;\;\; 
\frac{dC(t)}{dt} =  T [ R(-t) - R(t) ] 
\; ,
\end{equation} 
see Eqs.~(\ref{eq:chi-def}) and (\ref{eq:C-def}) 
for the definitions of $R$ and $C$.
Here and in what follows we set the Boltzmann constant to one, $k_B=1$. 
To write the second expression we used that $C(t)$ is an even function 
of $t$ (and $\dot C(t)$ is odd) and defined $\theta(0)\equiv 1/2$ (the 
same convention is used in the rest of the paper).  After Fourier
transforming the second expression becomes 
\begin{eqnarray} 
\omega C(\omega) = 2  T \im R(\o) \ , 
\label{eq:fdt-imaginary} 
\end{eqnarray} 
and the real part of $R(\o)$ is related to $\im R(\o)$ by the 
Kramers-Kr\"onig relation. 

The functions $g$ and $\nu$ are the integrated response and
correlation of the bath, respectively. 
As will be discussed in detail in section \ref{sec:II},
if the bath is itself in
equilibrium at a temperature $T$ they are related as
\begin{equation} 
\frac{1}{T} = \frac{2 \re g(\o) }{\nu(\omega)} 
\; . 
\end{equation} 
For an out of equilibrium bath we define the frequency dependent
temperature
\begin{equation} 
\frac{1}{T(\omega)} = \frac{2 \re g(\o) }{\nu(\omega) }
\; ,
\label{eq:Tomega-def}
\end{equation} 
and its inverse Fourier transform
\beq
T^{-1}(t)=
\Iint \frac{d\o}{2\p} \frac{2\re g(\o)}{\n(\o)} e^{-i\o t}=
\Iint dt' \n^{-1}(t-t') f(t')
\; .
\label{eq:Tt-def}
\eeq 
where
\beq \label{QQQ32}
\n^{-1}(t) =
\int \frac{d\o}{2\p} \frac{1}{\n(\o)} e^{-i\o t} \ .  
\eeq 
Note that both $f(t)$ and $T^{-1}(t)$ are even functions of $t$.
If the bath is in equilibrium at temperature $T$, $T^{-1}(t-t')=
\d(t-t')/T$. We shall assume throughout that $T^{-1}(t)$ goes to zero
fast enough for large $t$.

The main result of this paper is that the fluctuation relation for the
probability distribution function $\pi_\t(p)$ holds for long times $\t$:
\beq 
\t \sigma_+ \; p \sim \ln
\frac{\pi_\t(p)}{\pi_\t(-p)} \ , \eeq with \beq {\cal{S}}_\tau = \t \sigma_+
p = \int_{-\tau/2}^{\tau/2} dt \int_{-\tau/2}^{\tau/2} dt' \,
T^{-1}(t-t') \; \dot r_\a(t) \left[ - 
\frac{\partial V}{\partial r_\a}[\vec r(t')]
+ h_\a[\vec r(t')] \right] 
\; .
\eeq

\subsection{Large deviation function}

In the rest of this section we identify the entropy production rate
for Eq.~(\ref{eq:linear-t-harm}) following the procedure proposed by
Lebowitz and Spohn~\cite{LS99}.

The fluctuation relation is a symmetry property of the probability
distribution function ({\sc pdf}) of the entropy production rate
$\sigma_t$ that can be also expressed as a symmetry of the large
deviation function.  Calling $\sigma_+$ the average value of
$\sigma_t$, the {\sc pdf} of the variable $p$ defined in
Eq.~(\ref{pdef}) is defined as 
\beq 
\pi_\T(p)={\cal
P}\left[\frac{{\cal{S}}_\tau}{\T\sigma_+} = p \right] = \T\sigma_+ \langle
\, \delta ({\cal S}_\T - p \T \sigma_+ ) \, \rangle \; , 
\eeq 
where the
angular brackets denote an average over the realizations of the noise.
The large deviation function [normalized so that $\z(1)=0$ at the
maximum] is given by 
\beq
\label{zetapdef} 
\zeta(p)= 
\lim_{\T \rightarrow \infty} \T^{-1} \ln \big[ \pi_\T(p) / \p_\T(1) \big] 
\; .
\eeq  
It is easier to compute the characteristic function
\beq 
\label{phidef} 
\phi(\lambda) = - \lim_{\T \rightarrow \infty} \T^{-1}  
\ln \langle \exp[-\lambda \sigma_\T ] \rangle \ ;
\eeq 
the latter being the Legendre transform of $\z(p)$. Indeed,
\beq
e^{-\T \phi(\l)} = \langle e^{-\l \s_\T } \rangle = 
\frac{\int dp \, e^{\T [ \z(p) - \l p \s_+ ]}}{\int dp \, e^{\T \z(p)}} \sim
\frac{e^{\T \max_p [ \z(p)-\l p \s_+]}}{e^{\T \z(1)}}
\eeq
so that, recalling that $\z(1)=0$ by construction,
\beq
\phi(\l)=-\max_p \, [ \z(p)-\l p \s_+] \ .
\eeq
The inversion of the Legendre transform yields
\beq
\z(p)=\min_\l \, [ \l p \s_+ - \phi(\l) ] \ ,
\eeq
and it is 
easy to check that the fluctuation relation is equivalent 
to $\phi(\lambda)=\phi(1-\lambda)$. 

\subsection{Internal symmetries and the fluctuation relation}

Assume that there exists a map $I$ on the space of trajectories $r(t)$
such that $I^2 = 1$ and that the measure ${\cal D} r$ is inviariant
under $I$, i.e. ${\cal D} Ir = {\cal D}r$. Then, consider a segment of
trajectory $r(t)$, $t\in [-\t/2,\t/2]$ and define 
\beq 
{\cal S}_\t = -\ln
\frac{{\cal P}[Ir(t)]}{{\cal P}[r(t)]} \ , 
\eeq 
where ${\cal P}[r(t)]$
is the probability of observing $r(t)$, {\it in the stationary state}, for
$t\in [-\t/2,\t/2]$ irrespectively of what happens outside the
interval $[-\t/2,\t/2]$. It is easy to show that the {\sc pdf} of
${\cal S}_\t$ verifies the fluctuation theorem. Indeed 
\beq 
\langle e^{-\l
{\cal S}_\t} \rangle= \int {\cal D}r \, {\cal P}[r(t)] e^{-\l {\cal S}_\t} = \int
{\cal D}r \, {\cal P}[r(t)]^{1-\l} \, {\cal P}[Ir(t)]^\l = \int {\cal
D}r \, {\cal P}[Ir(t)]^{1-\l} \, {\cal P}[r(t)]^\l = \langle
e^{-(1-\l) {\cal S}_\t} \rangle \ .  
\eeq 
Thus, if the limit 
\beq
\label{chfunct} \phi(\l)=-\lim_{\t\rightarrow\io} \t^{-1} \ln\langle
e^{-\l {\cal S}_\t} \rangle 
\eeq 
exists, it verifies the relation $\phi(\l)
= \phi(1-\l)$ from which the fluctuation relation for the {\sc pdf}
of ${\cal S}_\t$ follows.  Lebowitz and Spohn showed that the limit
$\phi(\l)$ indeed exists for generic Markov processes and it is a
concave function of $\l$. Moreover they showed that ${\cal S}_\t$ can be
identified with the entropy production rate --over the time interval
$\t$-- in the stationary state up to boundary terms, {\it i.e.} terms that do
not grow with $\t$, if $I$ is chosen to be the time reversal,
$Ir(t)=r(-t)$.

\subsection{Entropy production rate}

We are interested in the explicit form of ${\cal{S}}_\tau$ for the equation of
motion (\ref{eq:linear-t-harm}) in the case in which
$g_{\a\b}(t)=\d_{\a\b} g(t)$ and $\vec h(t)=\vec h[\vec r(t)]$ is an
external nonconservative force that does not explicitly depend on
time: {\it e.g.}, in $d=2$, $\vec h = \a (-y,x)$.  Note that the
functions $\nu(t)$ and $g(t)$ are such that $\nu(t)=\nu(-t)$ while
$g(t)$ is proportional to $\th(t)$, and both decay sufficiently rapidly in
time. The probability distribution of the noise $\vec \r(t)$ is
\beq
\label{Prho} {\cal P}[\vec \r(t)] \propto \exp \left[ -\frac{1}{2}
\int dt dt' \, \r_\a(t) \n^{-1}(t-t') \r_\a(t') \right] 
\ , 
\eeq 
where
$\n^{-1}(t)$ is the operator inverse of $\n(t)$, see Eq.~(\ref{QQQ32}).
The
probability distribution of $\vec r(t)$ is obtained substituting $\vec
\r(t)$ obtained from Eq.~(\ref{eq:linear-t-harm}) in Eq.~(\ref{Prho}).
One has 
\beq\label{Pr}
\begin{split}
{\cal P}[\vec r(t)] \propto &\exp \left\{ - \frac{1}{2} \int dt dt' \, 
\left[
m \ddot r_\a(t) + \int dt'' g(t-t'') \dot r_\a(t'') 
+ \frac{\partial V}{\partial r_\a}[\vec r(t)] - h_\a[\vec r(t)]
\right] \right. 
\\ 
&\left. \times \; \n^{-1}(t-t') 
\left[
m \ddot r_\a(t') + \int dt''' g(t'-t''') \dot r_\a(t''') 
+ \frac{\partial V}{\partial r_\a}[\vec r(t')] - h_\a[\vec r(t')]
\right]
\right\} \ .
\end{split}
\eeq
After some algebra it is easy to see that
\beq\label{PIr}
\begin{split}
{\cal P}[\vec r(-t)] \propto &\exp \left\{ -\frac{1}{2}  \int dt dt' \, 
\left[
m \ddot r_\a(t) - \int dt'' g(t''-t) \dot r_\a(t'') 
+ \frac{\partial V}{\partial r_\a}[\vec r(t)] - h_\a[\vec r(t)]
\right] \right. 
\\ 
&\left. \times \; \n^{-1}(t-t') 
\left[
m \ddot r_\a(t') - \int dt''' g(t'''-t') \dot r_\a(t''') 
+ \frac{\partial V}{\partial r_\a}[\vec r(t')] - h_\a[\vec r(t')]
\right]
\right\} \ .
\end{split}
\eeq
To compute ${\cal{S}}_\tau$ we should consider the probability of a segment
of trajectory $[-\t/2,\t/2]$ and then send $\t$ to $\io$, neglecting
all boundary terms. As the functions $g(t)$ and $\nu(t)$ have short 
range, the trajectories $\vec r(t)$ decorrelate, say,  exponentially fast in
time and up to boundary contributions one can simply truncate the
integrals in ${\cal P}[\vec r(t)]$ in $t,t' \in [-\t/2,\t/2]$.

Let us now discuss the contributions to $ -\ln {\cal P}[\vec r(-t)] + 
\ln {\cal P}[\vec r(t)]$ that do not trivially vanish. 
One has:
\begin{itemize}
\item a ``kinetic'' term of the form
\beq\begin{split}
& \int dt dt' \, \left[
m \ddot r_\a(t) \n^{-1}(t-t') \int dt'' g(t''-t') \dot r_\a(t'')
+m \ddot r_\a(t) \n^{-1}(t-t') \int dt'' g(t'-t'') \dot r_\a(t'')\right] =\\
&= \int dt dt' \,
m \ddot r_\a(t) \n^{-1}(t-t') \int dt'' f(t'-t'') \dot r_\a(t'')=
 \int dt dt' \,
m \ddot r_\a(t) T^{-1}(t-t') \dot r_\a(t') \ .
\end{split}\eeq
If the bath is in equilibrium, this term trivially vanishes as it 
is the integral of the total derivative of the kinetic energy. 
But it also vanishes for a nonequilibrium bath. 
Indeed, by integrating by parts first in $t$ and then in $t'$, we find
\beq\label{zerokin}
\begin{split}
&\int dt dt' \, \ddot r_\a(t) T^{-1}(t-t') \dot r_\a(t') =
-\int dt dt' \, \dot r_\a(t) \frac{d}{dt} T^{-1}(t-t') \dot r_\a(t') = \\
&\int dt dt' \, \dot r_\a(t) \frac{d}{dt'} T^{-1}(t-t') \dot r_\a(t') =
-\int dt dt' \, \dot r_\a(t) T^{-1}(t-t') \ddot r_\a(t') = 0 \ .
\end{split}
\eeq 
where we used that $T^{-1}(t)$ is even and short ranged and we
neglected boundary terms.

\item a ``friction'' term of the form
\beq
\frac{1}{2}\int dt dt' dt'' dt''' \, \left[
 \dot r_\a(t'') g(t''-t) \n^{-1}(t-t') 
 g(t'''-t') \dot r_\a(t''') - 
 \dot r_\a(t'') g(t-t'') \n^{-1}(t-t') 
 g(t'-t''') \dot r_\a(t''')\right] \ .
\eeq
This term vanishes because the function
\beq
K(t''-t''')=\int dt dt' \, g(t''-t) \n^{-1}(t-t') 
 g(t'''-t') 
\eeq
is even in its argument as one can easily check.
\item a ``potential'' term of the form
\begin{eqnarray}
{\cal{S}}_\tau^V &=& - 
\int_{-\tau/2}^{\tau/2} dt 
\int_{-\tau/2}^{\tau/2} dt' 
\int_{-\tau/2}^{\tau/2} dt'' \, 
 f(t-t'') \dot r_\a(t'') 
 \n^{-1}(t-t') 
 \frac{\partial V}{\partial r_\a}[\vec r(t')] 
\nonumber\\
&=&
-
\int_{-\tau/2}^{\tau/2} dt 
\int_{-\tau/2}^{\tau/2} dt' 
\, T^{-1}(t-t')
 \dot r_\a(t) 
 \frac{\partial V}{\partial r_\a}[\vec r(t')] \ .
\end{eqnarray}
This term is related to the work of the conservative forces. If the
bath is in equilibrium, it vanishes being the total derivative of the
potential energy. It vanishes also for a harmonic potential $V(\vec
r)=\frac{1}{2} k r^2$ because $\frac{\partial V}{\partial r_\a}[\vec r(t)]=k
r_\a(t)$ and one can use the same trick used in Eq.~(\ref{zerokin}).
It does not vanish in general. 

\item a ``dissipative'' term 
\beq 
{\cal{S}}_\tau^{diss}=
\int_{-\tau/2}^{\tau/2} dt \int_{-\tau/2}^{\tau/2} dt' \, T^{-1}(t-t')
\dot r_\a(t) h_\a[\vec r(t')] 
\; .  
\eeq 
This term is related to the
work of the dissipative forces.  If the bath is in equilibrium at
temperature $T$, this is exactly the work of the dissipative forces
divided by the temperature of the bath. Otherwise, the work done at
frequency $\o$ is weighted by the effective temperature at the same
frequency.
\end{itemize}

The expression of the total entropy production over the interval
$[-\tau/2,\tau/2]$ is then
\begin{eqnarray}
{\cal{S}}_\tau &=& {\cal{S}}_\tau^V+{\cal{S}}_\tau^{diss} =  
\int_{-\tau/2}^{\tau/2} dt \int_{-\tau/2}^{\tau/2} dt' \, T^{-1}(t-t')
 \dot r_\a(t) \left[ - \frac{\partial V}{\partial r_\a}[\vec r(t')] + 
h_\a[\vec r(t')] \right] 
\nonumber\\
&=& 
 \int_{-\tau/2}^{\tau/2} dt \int_{-\tau/2}^{\tau/2} dt' \, T^{-1}(t-t')
 \dot r_\a(t) F_\a(t') 
\; , 
\label{eq:generic-result}
\end{eqnarray}
where $F_\a(t) = h_\a[\vec r(t)]- \frac{\partial V}{\partial r_\a}[\vec r(t)]$ 
is the total deterministic force acting on the particle at time $t$. 

The latter expression can be rewritten as
\beq
{\cal{S}}_\tau \sim \int_{-\t/2}^{\t/2} dt \; \s_t =  
\int_{-\t/2}^{\t/2} dt  \; (\s_t^V + \s_t^{diss})
\eeq
defining a entropy production rates $\s_t$, $\s_t^V$ and $\s_t^{diss}$  
(modulo subdominant terms in the large $\t$ limit) as
\beq \label{EPR} \begin{split}
&\s_t = \s^V_t + \sigma^{diss}_t = \int_{-\io}^t dt' \, T^{-1}(t-t')
\big[ \dot r_\a(t) F_\a(t') + \dot r_\a(t') F_\a(t) \big] \ , \\%
&\s^V_t =- \int_{-\io}^t dt' \, T^{-1}(t-t')
\left[ \dot r_\a(t)\frac{\partial V}{\partial r_\a}[\vec r(t')]  + 
\dot r_\a(t') \frac{\partial V}{\partial r_\a}[\vec r(t)] \right] \ , \\
&\sigma^{diss}_t = \int_{-\io}^t dt' \, T^{-1}(t-t')
\big[ \dot r_\a(t) h_\a[\vec r(t')] + \dot r_\a(t') h_\a[\vec r(t)] \big] \ .
\end{split} \eeq

We recall that if the bath is in equilibrium this expression
reduces to the work done by the nonconservative forces divided by the
temperature of the bath, as expected. If the bath is not in
equilibrium, but the potential is harmonic, only the contribution
$\sigma^{diss}_\t$ of the nonconservative force has to be taken into
account.  The reason why the work of the conservative forces produces
entropy if the bath is out of equilibrium {\it and} the interaction is
nonlinear is that the nonlinear interaction couples modes of different
frequency which are at different temperature, thus producing an energy
flow between these modes; this energy flow is related to the entropy
production.

It is also important to remark that boundary terms, that are usually
neglected, can have dramatic effects on the large fluctuations of
${\cal S}_\t$ even for $\t \to \io$, as pointed out by Van Zon and Cohen
\cite{VzC04}. This happens if the {\sc pdf} of the boundary term has
exponential tails. Thus, boundary terms {\it cannot be always
neglected}, at least for very large values of $p$.  A good empirical
prescription to remove boundary contributions is the following: in
equilibrium $\s_t$ must be a total derivative as no dissipation is
present.  So, removing a total derivative (a boundary term) one can
define ${\cal S}_\t$ in such a way that it vanishes identically in
equilibrium. This definition turns out to be the one that verifies the
fluctuation relation for all $|p|<p^*$, $p^*$ being the maximum
allowed value of $p$ for $\t \to \io$~\cite{VzC04,BGGZ05}. We shall
discuss this point in more detail later.


\section{A driven  particle in a harmonic potential} 
\label{sec:II}  
 
In this and the next Section we discuss some examples on which we test
the fluctuation relation for ${\cal S}_\t$ defined in
(\ref{eq:generic-result}).  We first consider
the simplest fully analytically solvable case in which there are no
applied forces and the potential is quadratic.  We derive the
fluctuation-dissipation relation between induced and spontaneous
fluctuations in the position of the particle and we relate it to the
time-dependent temperature of the bath defined in Eq.~(\ref{eq:Tt-def}).
We then show that in this simple case the fluctuation relation for ${\cal S}_\t$
in (\ref{eq:generic-result}) reduces to the usual one with the temperature
of the bath.
 
\subsection{The fluctuation-dissipation relation}
\label{sec:FDT}

In the harmonic Brownian particle problem with no other  
applied external forces the dynamics of 
different spatial components are not coupled. Thus, without loss of  
generality, we henceforth focus on $d=1$.  
In Fourier space, the Langevin equation reads 
\begin{equation} 
-m \omega^2 x(\omega)  -i\omega g(\omega) x(\omega) =  
-k x(\omega) + \rho(\omega) 
\label{eq:linear-omega-harm} 
\end{equation} 
with the noise-noise correlation 
\begin{equation} 
\langle \, \rho(\omega) \rho(\omega') \, \rangle = 
2\pi \delta(\omega+\omega') \nu(\omega) 
\; . 
\end{equation} 
The linear equation (\ref{eq:linear-omega-harm}) is solved by 
\begin{equation} \label{eq17}
x(\omega) = G(\omega) \rho(\omega) \ ,
\;\;\;\;\;\;\;\;\;\;\;\;\; 
G(\omega) \equiv \frac{1}{-m\omega^2 -i \omega g(\omega) + k}  
\;,  
\end{equation} 
and one finds the correlations 
\beq
\begin{split}
& \langle \, x(\omega) x(\omega') \, \rangle = 
G(\omega) G(-\omega) 2\pi \delta(\omega+\omega') \nu(\omega) \ ,
\\ 
& \langle \, x(\omega) \rho(\omega') \, \rangle = 
G(\omega)  
2\pi \delta(\omega+\omega') \nu(\omega)  
\; . 
\label{eq:resp-omega-harm} 
\end{split}
\eeq 
Note that $G(\omega) G(-\omega)=|G(\omega)|^2$; then  
\begin{eqnarray} 
&& \langle \, x(\omega) x(\omega') \, \rangle = 
C(\omega) 2\pi \delta(\omega+\omega') 
\;\;\;\;\;\;\;\;\;\; 
\mbox{with} 
\;\;\;\;\;\;\;\;\;\; 
C(\omega) \equiv |G(\omega)|^2 \nu(\omega)  
\label{eq:corr-omega-harm2} 
\; . 
\end{eqnarray} 
In a problem solved by  
\begin{equation} 
x(t) = \Iint dt' \; G(t-t') [\rho(t')+h(t')] + \mbox{IC} 
\; , 
\end{equation} 
where IC are terms related to the initial conditions, 
the time-dependent linear response is  
\begin{equation} 
R(t-t') \equiv   
\left. \frac{\delta \langle x(t)\rangle}{\delta h(t')}\right|_{h=0} 
= G(t-t')  
\; , 
\end{equation} 
and 
\beq 
R(\o) = \Iint dt \ e^{i\o t} R(t) = G(\o) \ . 
\eeq 
Note that the response function is related to the correlation  
$\langle \, x(t) \r(t') \, \rangle$ by Eq.~(\ref{eq:resp-omega-harm}): 
\begin{equation} 
2\pi \delta(\omega+\omega') R(\omega) \nu(\omega) =  
\langle \, x(\omega) \rho(\omega') \, \rangle  
\; . 
\end{equation} 
 
Now, we can check under which conditions on the characteristics of the
bath [$g(t-t')$ and $\nu(t-t')$] the fluctuation-dissipation theorem
(for the particle) holds and, when it does not hold, which is the
generic form that the relation between the linear response and
correlation might take in this simple quadratic model.  Eq.
(\ref{eq17}) implies~\cite{comment2}
\begin{equation} 
\mbox{Im} R(\omega) =  \mbox{Im} G(\omega) =  
\omega \; \mbox{Re}g(\omega) \; |G(\omega)|^2 \ ,
\label{eq:fdt-generic-bath0} 
\end{equation} 
and then using equation~(\ref{eq:corr-omega-harm2})
\begin{equation} 
\frac{\omega C(\omega)}{2 \mbox{Im} R(\omega)}= 
\frac{\nu(\omega)}{2\mbox{Re}g(\omega)}
= T(\omega) 
\label{eq:fdt-generic-bath} 
\; . 
\end{equation} 
We see that the fluctuation-dissipation theorem holds only if this
ratio is equal to $ T$, see Eq.~(\ref{eq:fdt-imaginary}). Otherwise,
the relation between linear response and correlation of the particle
is given by the frequency-dependent temperature of the bath,
$T(\omega)$, defined in (\ref{eq:Tomega-def}).  The measure of the 
modification of the fluctuation-dissipation theorem given in 
(\ref{eq:fdt-generic-bath}) is the  
{\it effective temperature} of the system. The use of this name 
has been justified within a number of
models with slow dynamics and a separation of
time-scales~\cite{Cukupe,Cukuja} but it might not hold in complete
generality~\cite{proviso}.
It is important to remark that the effective temperature in the frequency
domain, $T(\o)$ is {\it not} equal in 
general to the Fourier transform of the effective temperature 
$T(t)$ --defined in the introduction-- 
which is the ratio between the noise-noise correlation and 
memory function in the time domain.

Let us now discuss some environments that we shall 
use in the rest of the paper. 
 
\subsubsection{Equilibrated environments} 
 
For any environment such that the right-hand-side in  
Eq.~(\ref{eq:fdt-generic-bath})  
equals $ T$  the fluctuation-dissipation theorem holds. 
In the time domain, this condition reads 
\beq 
\label{FDTbathTD} 
 T g(t) = \theta(t) \nu(t) \; , \;\;\;\;\;\;\;\;\;\; \nu(t) = T [g(t)
+ g(-t)] = T g(|t|) \ .  
\eeq 
In particular, this is satisfied by a
white noise for which $\nu(t) = 2 T \gamma \delta(t)$ and $g(t) = 2
\gamma \delta(t) \theta(t)$ (remember that $\theta(0) \equiv 1/2$).
The fluctuation-dissipation theorem also holds for any colored noise
-- with a retarded memory kernel $g$ and noise-noise correlation $\nu$
-- such that the ratio between Re$g(\omega)$ and $\nu(\omega)$ equals
$(2 T)^{-1}$. This requirement applies to any equilibrated bath.

\subsubsection{Nonequilibrium environments} 
\label{subsec:noneq-baths} 
 
Instead, for any other generic environment, 
the left-hand-side in  
Eq.~(\ref{eq:fdt-generic-bath}) yields a non-trivial and,  
in general model-dependent, result for the effective temperature. 
 
A special case that we shall study in Appendix~B is the one of   
an ensemble of $N$ equilibrated baths with different  
characteristic times and at different temperatures. In  
this case, the noise $\vec\r$ in Eq.~(\ref{eq:linear-t-harm}) is 
the sum of $N$ independent noises, 
\beq 
\vec\r = \sum_{i=1}^N \vec\r_i \ , 
\hspace{20pt} 
\langle  \, 
\r_{i\a}(t) \r_{j\b}(t') \, \rangle 
= 
\d_{\a\b} \d_{ij}  T_i \nu_i(t-t') \ , 
\eeq 
and the friction kernel is given by 
\beq 
\label{eq:frict-sumN} 
g(t-t')=\sum_{i=1}^N g_i(t-t')  
\ .   
\eeq  
We have extracted the 
temperature $T_i$ from the definition of $\nu_i(t)$ in order to 
simplify several expressions. As 
the $\vec\r_i$ are independent Gaussian variables, $\vec\r = 
\sum_i \vec\r_i$ is still a Gaussian variable with zero mean and 
correlation \beq 
\label{eq:corr-sumN} 
\langle \r_\a(t) \r_\b(t') \rangle = \d_{\a\b} \sum_i T_i \nu_i(t-t')
\ .  \eeq Thus, in the Gaussian case {\it the $N$ equilibrated baths
are equivalent to a single nonequilibrium bath} with correlation given
by Eq.~(\ref{eq:corr-sumN}) and friction kernel given by
Eq.~(\ref{eq:frict-sumN}).  In frequency space we have
\begin{eqnarray} 
g(\omega) =\sum_{i=1}^N g_i(\omega)  
\; ,  
\;\;\;\;\;\;\;\;\;\;\;\;\;\;\;\;\;\; 
\nu(\omega) =\sum_{i=1}^N  T_i \nu_i(\omega) 
\; ,   
\end{eqnarray} 
with  
\begin{eqnarray} 
\label{FDTbathFD} 
\nu_i(\omega)= 2\mbox{Re} g_i(\omega) 
\; , 
\end{eqnarray} 
as each bath is equilibrated at temperature $T_i$. 
The frequency-dependent  temperature is then given by 
\beq 
\label{TeffNbaths} 
T(\o) = \frac{\sum_{i=1}^N T_i \nu_i(\o)}{\sum_{i=1}^N
\nu_i(\o)} \ .  
\eeq 
Note that if the functions $\nu_i(\o)$ are chosen
to be peaked around a frequency $\o_i$, by choosing suitable values of 
$\o_i$ and $T_i$ one can approximate a single nonequilibrium bath with
$N$ baths equilibrated at different temperatures.

\subsection{Large deviation function}
 
We now compute the large deviation function in the
harmonic case, $V(\vec r) = \frac{1}{2} k r^2$.  In this case $\s^V_\t$
is a total derivative and only the term $\sigma^{diss}_\t$, related to
the nonconservative forces, is relevant. We show that the
characteristic function $\phi(\l)$ of $\sigma^{diss}_\t$ exists, is a
convex function of $\l$ and verifies the fluctuation relation
$\phi(\l)=\phi(1-\l)$.

\subsubsection{Equilibrium bath}

As a first illustrative example we consider the case of an equilibrium
white bath.  The model we study is a two dimensional harmonic
oscillator with potential energy $V(x,y)=\frac{k}{2}(x^2+y^2)$ coupled
to a simple white bath in equilibrium at temperature $T$, and driven
out of equilibrium by the nonconservative force $\vec h=\alpha
(-y,x)$.  The equations of motion are \beq
\begin{split} 
&m \ddot x_t + \gamma \dot x_t = -k x_t -\alpha y_t + \xi_t \ , \\ 
&m \ddot y_t + \gamma \dot y_t = -k y_t +\alpha x_t + \eta_t \ , 
\end{split} 
\eeq 
where $\xi_t$ and $\eta_t$ are independent Gaussian white noises with  
variance 
$\langle \, \xi_t \xi_0 \, \rangle = \langle \, \eta_t \eta_0 \, \rangle = 
2\gamma T \delta(t)$. 
The memory friction kernels $g_{\a\b}(t-s)$ 
are simply $\delta_{\a\b} g(t-s)= 
2 \delta_{\a\b} \gamma \delta(t-s) \theta(t-s)$ in this case, with  
$\gamma$ the friction coefficient. 
 
Defining the complex variable $z_t=(x_t+iy_t)/\sqrt{2}$ and the noise  
$\r_t=(\xi_t+i\eta_t)/\sqrt{2}$ the equations of motion can be written as 
\beq 
\label{whitemotion} 
m \ddot z_t + \gamma \dot z_t = -\kappa z_t +\rho_t \ , 
\eeq 
where $\kappa = k -i\alpha$,  
$\langle \rho_t \rho_0 \rangle 
=  \langle \bar\rho_t \bar\rho_0 \rangle = 0$ and 
$\langle \rho_t \bar\rho_0 \rangle = 2 \gamma T \delta(t)$. 
The complex noise $\rho_t$ has a Gaussian {\sc pdf}: 
\beq 
{\cal P}[\rho_t] \propto \exp \left[ -\frac{1}{2 \gamma T}  
\int_{-\infty}^\infty dt \ \rho_t \bar \rho_t \right] 
\; . 
\label{eq:Gaussian-pdf-rho} 
\eeq 
The energy of the oscillator is $H= m \dot z \dot{\bar z} + k z \bar 
z$, and its time derivative is given by  
\beq  
\frac{dH}{dt} = 2 m \re 
\dot z_t \ddot{\bar z}_t + 2 k \re z_t \dot{\bar z}_t =2 \alpha \im \dot 
z_t\bar z_t -
2 \gamma \dot z_t \dot{\bar z}_t + 2 \re \dot z_t \bar \rho_t = 
 W_t - \widetilde W_t \ ,  
\eeq  
where $W_t = 2 \alpha 
\im \dot z_t \bar z_t = \alpha (x_t \dot y_t - y_t \dot x_t)$ is the 
power injected by the driving force and $\widetilde W_t = 2 \gamma 
\dot z_t \dot{\bar z}_t - 2 \re \dot z_t \bar \rho_t$ is the power 
extracted by the thermostat (henceforth we choose the sign of the 
power in order to have positive average).  

The entropy production rate (\ref{EPR}) 
reduces, as expected, to the
injected power divided by the temperature, $\sigma_t = \b W_t$, where
$\b=1/T$ (one could also consider the entropy production of the bath,
$\widetilde \sigma_t = \b \widetilde W_t$; for completeness 
we discuss it in Appendix~\ref{app:B}).
 
We want to compute the probability distribution function ({\sc pdf})
of the entropy production rate $\sigma_t = \b W_t$.  The average value
of $\sigma_t$ is in this case given by $\sigma_+ = 2 \alpha^2 /(\g
k)$.  From Eq.~(\ref{eq:generic-result}) 
we can rewrite the total entropy production
${\cal{S}}_\tau $ in terms of the
complex variable $z_t$: 
\beq {\cal{S}}_\tau = \int_{-\T/2}^{\T/2} dt \
\sigma_t = 2 \alpha \beta \; \im \int_{-\T/2}^{\T/2} dt \ \dot z_t
\bar z_t \ .  
\eeq 
As already discussed, it is easier to compute the
characteristic function $\phi(\lambda)$, Eq.~(\ref{chfunct}), in terms
of which the fluctuation relation reads $\phi(\l)=\phi(1-\l)$.  To
leading order in $\T$ we can neglect all the boundary terms in the
integrals. After integrating by parts we have \beq {\cal{S}}_\tau  = 2
\alpha \beta i \int_{-\T/2}^{\T/2} dt \ z_t \dot{\bar z}_t \ .  \eeq
In terms of the {\sc pdf} of the noise (\ref{eq:Gaussian-pdf-rho}) we
obtain \beq
\label{AAA1} 
\langle \exp[- \lambda {\cal{S}}_\tau  ] \rangle = {\cal N}^{-1} 
\int d{\cal P}[\rho_t] \ \exp \left[ - \frac{2i \alpha \lambda}{T} 
\int_{-\T/2}^{\T/2} dt \ z_t \dot{\bar z}_t \right]  
\ , 
\eeq 
and the normalization factor ${\cal N} = \int d{\cal P}[\rho_t]$  
is simply given by the numerator calculated at $\lambda =0$. 
 
To leading order in $\T$ the function $\phi(\l)$ should not depend 
on the boundary conditions in Eq.~(\ref{AAA1}). 
Thus, we impose periodic boundary conditions, 
$z(\T/2)=z(-\T/2)$ and $\dot z(\T/2)=\dot z(-\T/2)$, 
and we expand $z_t$ in a Fourier series, 
\beq 
\label{AAA2} 
z_t =  
\frac{\Delta \omega}{2 \pi} 
\sum_{n=-\infty}^{\infty} z_n \ e^{-i \omega_n t}  
\ , 
\eeq 
where $\Delta \omega= 2 \pi /\T$ and $\omega_n = n \Delta \omega$. 
For $\T \rightarrow \io$ 
\beq 
z_t = 
\int_{-\infty}^\infty \frac{d\omega}{2\pi} \ e^{-i\omega t} z_\omega \ , 
\hspace{30pt} z_\omega = \int_{-\infty}^\infty dt \ e^{i\omega t} z_t
\ , 
\eeq 
and the equations of motion become 
\beq 
z_\omega =
\frac{\rho_\omega}{ - \omega^2 m +\kappa - i\omega \gamma } \equiv
\frac{\rho_\omega}{D(\o)} 
\ .  
\eeq 
Note that in the limit $\alpha=0$
the Green function $G(\a,\o) = 1/D(\o)$ reduces to the one used above
 to compute the violation of the
fluctuation-dissipation theorem induced by a nonequilibrium bath.  The
distribution of the noise is given by 
\beq
\label{AAA3} 
{\cal P}[\r_\o] = \exp \left[ -\frac{1}{2 \gamma T}  
\int_{-\infty}^\infty \frac{d\omega}{2\pi} \  
\rho_\omega \bar \rho_\omega \right] 
\sim  \exp \left[ -\frac{1}{2 \gamma T}  
\frac{\D\omega}{2\pi}\sum_{n=-\infty}^\infty \rho_n \bar \rho_n \right]  
\ . 
\eeq 
Substituting Eqs.~(\ref{AAA2}) 
and~(\ref{AAA3}) into Eq.~(\ref{AAA1}) we get 
\beq 
\label{AAA4} 
\langle \exp[- \lambda {\cal{S}}_\tau  ] \rangle = 
 {\cal N}^{-1} \int d\r_n 
\exp \left[ -  \frac{\Delta \omega}{2\pi}  
\sum_{n=-\infty}^{\infty} \left( 
\frac{|\r_n|^2}{2 \g T} -  
\frac{2 \a \l \o_n |\r_n|^2}{T |D(\o_n)|^2} \right)\right] 
=  
\prod_{n=-\io}^\io  
\left[ 1 - \frac{4 \g \a \l \o_n}{|D(\o_n)|^2} \right]^{-1}  
\eeq 
and using Eq.~(\ref{phidef}) 
\beq 
\label{phi1bagnoFourier} 
\phi(\lambda) =  \lim_{\T \rightarrow \infty}  
\frac{1}{\T} \sum_{n=-\infty}^{\infty}  
\ln \left[ 1 - \frac{4 \g \a \l \o_n}{|D(\o_n)|^2} \right] = 
\int_{-\infty}^\infty \frac{d\omega}{2 \pi} 
\ln \left[ 1 - \frac{ 4 \g \a \l \o} 
{ |D(\o)|^2 } \right] \ . 
\eeq 
To show that $\phi$ verifies $\phi(\lambda)=\phi(1-\lambda)$ and  
hence the fluctuation theorem, note that 
\beq 
\label{AAA5} 
\phi(\lambda)-\phi(1-\lambda)=\int_{-\infty}^\infty \frac{d\omega}{2 
\pi} \ln \left[ \frac{ |D(\o)|^2 - 4 \g \a \l \o} { |D(\o)|^2 - 4 \g 
\a (1-\l) \o } \right]  
\eeq  
and, as $|D(\o)|^2 - 4 \a \g \o = 
|D(-\o)|^2$, the integrand is an odd function of $\o$ and the integral 
vanishes by symmetry.  In Appendix~\ref{app:A} we show that the same 
result is obtained if one uses Dirichlet boundary 
conditions (at least for $m=0$, where the computation is feasible); 
this result supports the approximations made when neglecting all the 
boundary terms in the exponential. Moreover, in the case $m=0$ the 
large deviation function $\z(p)$ can be explicitly calculated; 
defining $\t_0 = \g/k$ and $\s_0 = \s_+ \t_0/2 = \a^2/k^2$, we obtain 
\beq 
\label{zetap} 
\z(p)=\t_0^{-1} \left[ 1 + p \s_0 - \sqrt{(1+\s_0)(1+p^2 \s_0)} 
\right] \ .   
\eeq  
Thus, for $\t \rightarrow \infty$ the {\sc pdf} of $p$ has the form 
\beq 
\pi_\t(p) \propto \exp \left[ \frac{\t}{\t_0} \tilde\z(p,\s_0)\right] \ . 
\eeq 
Note that $\t_0$ is the decay time of the correlation function of $z_t$  
[\ie $\langle z_t \bar z_0 \rangle \propto \exp(-t/\t_0)$] and $\s_0$ is 
the average entropy production over a time $\t_0 / 2$. 
Thus, $\t_0$ is the natural time unit of the problem (as expected); 
remarkably, the function $\tilde\z=\t_0 \z$ depends only on 
$\s_0$ and not on the details of the model. 
It would be interesting to see whether the same scaling holds for more  
realistic models. 
 
In summary, we found that for all driving forces, {\it i.e.} all
values of $\alpha$, the fluctuation theorem holds for the entropy
production rate (\ref{eq:generic-result}).  
For a white equilibrium bath this result
has already been obtained in~\cite{Ku98}.  The temperature entering
the fluctuation theorem is the one of the equilibrated environment
with which the system is in contact, although it is not in equilibrium
with it, when the force is applied.
 
Let us also stress that one can easily check that the
fluctuation-dissipation relation holds in the absence of the drive
(see Sect.~\ref{sec:FDT}) but it is strongly violated when the system is
taken out of equilibrium by the external force.
 
\subsubsection{Non-equilibrium bath} 
\label{sec:non-eq}
 
We now generalize the calculation to the case of a generic 
nonequilibrium bath; the equation of motion becomes
\beq 
\label{motioncomplex} 
m \ddot z_t + \int_{-\infty}^\infty dt' \ g(t-t') \dot z_{t'} = -\kappa z_t 
+ \rho_t \ ,  
\eeq  
where as before $\kappa=k-i\a$ and $\langle \r_t \bar 
\r_0 \rangle = \nu(t)$. The functions $\nu(t)$ and $g(t)$ are now arbitrary 
(apart from the condition $g(t)=0$ for $t<0$), hence they do not 
satisfy, in general, Eq.~(\ref{FDTbathTD}). 
Note that Eq.~(\ref{motioncomplex}) provides 
a model for the dynamics of a confined Brownian particle in an out of 
equilibrium medium \cite{Po04,AG04,PM04}.  

The dissipated power is given by  
\beq  
\frac{dH}{dt} 
= 
2\alpha \; \im \dot z_t \bar z_t -2\re \int_{-\infty}^\infty dt' \ g(t-t') 
\dot z_t \dot{\bar z}_{t'} + 2 
\re \dot z_t \bar \rho_t = W_t - \widetilde W_t \; ,  
\eeq  
where as in the previous case $W_t = 2\alpha \im \dot z_t \bar 
z_t $ is the power injected by the external force and $\widetilde 
W_t = 2\re \int_{-\infty}^\infty dt' \ g(t-t') \dot z_t \dot{\bar 
z}_{t'} - 2 \re \dot z_t \bar \rho_t$ is the power extracted by the 
bath. 

For the harmonic model $\s^V_\t$ is a boundary term and 
Eq.~(\ref{eq:generic-result}) gives
\beq
\label{sigma3} 
{\cal{S}}_\tau ^{diss} = -2 \a  \frac{\D\o}{2\p} 
\sum_{n=-\io}^\io \frac{\o_n |z_n|^2}{T(\o_n)} = 
2 \alpha i \int_{-\T/2}^{\T/2} dt  
\int_{-\T/2}^{\T/2} dt' \ T^{-1}(t-t') z_t \dot{\bar z}_{t'} \ .
\eeq 
Note that the last equality holds neglecting boundary terms. 

Let us now compute $\phi_{diss}(\l)$.
The computation is straightforward following the strategy of 
section~\ref{sec:II}. 
In Fourier space, Eq.~(\ref{motioncomplex}) reads  
\beq  
z_\o=\frac{\r_\o}{-m 
\o^2 + \k - i\o g(\o)}=\frac{\r_\o}{D(\o)} \ .   
\eeq  
The probability distribution of $\r_\o$ is  
\beq  
{\cal P}[\r_\o] = \exp \left[ -
\int_{-\io}^\io \frac{d\o}{2\p} \frac{|\r_\o|^2}{\nu(\o)}\right] 
\ . 
\eeq  
Thus [see Eqs.~(\ref{AAA3}) and (\ref{AAA4})],  
\beq  \label{QQQ27}
\langle \exp[- 
\lambda {\cal{S}}_\tau^{diss}  ] \rangle 
= {\cal N}^{-1} \int d\r_n \exp \left[ - 
\frac{\Delta \omega}{2\pi} \sum_{n=-\infty}^{\infty} \left( 
\frac{|\r_n|^2}{\nu(\omega_n)} - \frac{2 \a \l \o_n |\r_n|^2}{T(\o_n)
|D(\o_n)|^2} \right)\right] =  
\prod_{n=-\io}^\io \left[ 1 - \frac{2 \a 
\l \o_n \nu(\o_n)}{T(\o_n) |D(\o_n)|^2} \right]^{-1} \ ,  
\eeq  
and using the definition of $T(\o)$ given by Eq.~(\ref{eq:Tomega-def})
\beq
\label{phi_eff}
\phi_{diss}(\l) = \int_{-\io}^\io \frac{d\o}{2\p} \ln \left[ 1 - 
\frac{4 \a \l \o \re g(\o)}{|D(\o)|^2} \right] \ .   
\eeq  
It is easy to 
prove that $|D(\o)|^2 - 4 \a \o \re g(\o) = |D(-\o)|^2$. Using 
now the same trick employed in Eq.~(\ref{AAA5}), one shows that 
$\phi_{diss}(\l)=\phi_{diss}(1-\l)$. 

An alternative definition of entropy production rate in which one
assumes that it is proportional to the power injected by the external
drive, $\su_t = \Th^{-1} W_t$, via a parameter $\Th$ which has the
dimension of a temperature, has been often used in the
literature~\cite{GGK01,CGHLP03,FM04,Ga04}.  With this definition, the
total entropy production over a time $\T$ is given by (neglecting
boundary terms) 
\beq
\label{sigma1} 
{\cal{S}}_\tau^\Theta 
= 
\frac{2 \alpha i}{\Th} \int_{-\T/2}^{\T/2} dt \ z_t \dot{\bar
z}_t = -2 \a \frac{\D\o}{2\p}\sum_{n=-\io}^\io \frac{\o_n
|z_n|^2}{\Th} 
\; ,
\eeq 
\ie $T(\o)$ is replaced by a constant
$\Th$ that is taken as a free parameter that
one adjusts in such a way that the {\sc pdf} of ${\cal S}^\Th_\tau$ 
is as a close as
possible to verify a fluctuation relation~\cite{GGK01,CGHLP03,FM04}.
Substituting $T(\o)$ with a constant $\Th$ in Eq.~(\ref{QQQ27}) one
obtains
\beq 
\label{phi_theta}
\phi_\Th(\l) = \int_{-\io}^\io \frac{d\o}{2\p} \ln \left[ 1 - 
\frac{2 \a \l \o \nu(\o)}{\Th |D(\o)|^2} \right] 
\ .   
\eeq  
However, it is not possible to find a value of $\Th$
such that $\phi_\Th(\l)$ satisfies the fluctuation theorem for the
harmonic problem in contact with a generic bath.  
We shall show in section~\ref{sec:approximate} that the use of a 
single parameter $\Theta$ constitutes a rather good approximation 
when the dynamics of the particle occurs on a single time scale.
 
In conclusion, the fluctuation theorem is satisfied when the entropy
production rate is defined using the power injected by the external
drive with the temperature of the environment defined as in 
(\ref{eq:Tomega-def}).

In Sect.~\ref{subsec:noneq-baths} we also introduced a complex bath
made of many equilibrated baths at different temperatures, eventually
acting on different time scales.  In Appendix~\ref{app:III} we prove
that, as expected, the {\sc pdf} of ${\cal S}_\tau^{diss}$ defined in
Eq.~(\ref{sigma3}) also verifies the fluctuation theorem in this case
-- while the {\sc pdf} of ${\cal S}_\tau^\Th$ does not.  For such a
multiple bath one can also consider the entropy production of the
baths, defined as the power extracted by each bath divided by the
corresponding temperature. This quantity is of interest if one could
identify the different thermal baths with which the system is in
contact; clearly, this is not possible in glassy systems where the
effective temperature is self-generated.  Nevertheless, the study of
systems of particles coupled to many baths at different temperature is
of interest in the study of heat conduction.  In Appendix~\ref{app:B}
we prove that the entropy production rate of the baths verifies the
fluctuation theorem, at least for $|p|\leq 1$ 
(see also~\cite{VzC04,BGGZ05}).

 
\section{A driven particle in an anharmonic potential: numerical results} 
\label{sec:V}
 
In this section we investigate numerically Eq.~(\ref{motioncomplex})
for a particular choice of the nonequilibrium bath and in presence of
a linear and nonlinear interaction.  In the linear case, we find that
the numerical results reproduce the analytical results of the previous
section. This finding confirms that the boundary terms we neglected in
the analytical computation are indeed irrelevant.  In the nonlinear
case, we find that  the fluctuation relation seems to be satisfied
quite well for
${\cal{S}}_\tau^{diss}$, although strictly speaking it only holds for
${\cal{S}}_\tau^{diss}+{\cal{S}}_\tau^{V}$. We shall discuss the reason 
for this below.

We consider the simplest non trivial case, 
in which a massless Brownian particle is coupled to two equilibrated 
baths: a white (or fast) bath at temperature $T_f$ and a colored (or 
slow) bath with exponential correlation at temperature $T_s$. This
model has been studied in detail in \cite{Cukuja} and is relevant for the 
description of glassy dynamics when the time scales of the two baths 
are well separated, as will be discussed in Sect.~\ref{sec:VI}.
The equations of motion are given by 
Eq.~(\ref{motioncomplex}) with $g(t)=g_f(t)+g_s(t)$, 
$g_f(t)=\g_f \d(t)$ and 
$g_s(t)=\th(t) \frac{\g_s}{\t_s}e^{-\frac{t}{\t_s}}$, or, equivalently, 
$g_f(\o)=\g_f$ and
$g_s(\o)=\g_s/(1-i\o\t_s)$. We use a 
generic rotationally invariant 
potential $V(x,y)={\cal V}\left(\frac{x^2+y^2}{2}\right)={\cal V}(|z|^2)$.
The noise is the sum of a fast and a slow component.
Then Eq.~(\ref{motioncomplex}) becomes  
\beq 
\label{motion2baths} 
\g_f \dot z_t + \frac{\g_s}{\t_s}  
\int_{-\io}^t dt' \ e^{-\frac{t-t'}{\t_s}} \dot z_{t'} = 
- z_t {\cal V}'(|z_t|^2) + i\a z_t + \r^f_{t} + \r^s_{t} 
\ , 
\eeq 
where $\langle \r^f_{t} \r^f_{t'} \rangle 
= 2 \g_f T_f \delta(t-t')$, 
$\langle \r^s_{t} \r^s_{t'} \rangle = 
\frac{T_s \g_s}{\t_s} e^{-|t-t'|/\t_s}$ 
and ${\cal V}'(x)$ is the derivative of ${\cal V}(x)$ wtih respect to $x$. 
It is convenient to rewrite this equation as
\begin{eqnarray}
\label{riscritte}
\left\{
\begin{array}{l}
\dot u_t = 
- \frac{u_t - \y_t}{\t_s} + \frac{\g_s z_t}{\t_s^2} 
\ , 
\\ 
\g_f \dot z_t 
= 
- z_t  {\cal V}'(|z_t|^2) + i\a z_t + \r^f_{t} + u_t - 
\frac{\g_s z_t}{\t_s} 
\ , 
\end{array}
\right.
\end{eqnarray}
where we introduced the auxiliary variable $u_t$ and $\y_t$ is a
white noise with correlation $\langle \y_t \bar\y_{t'} \rangle = 2
\g_s T_s \delta(t-t')$.  The power injected by the external force is,
as usual, $W_t=2\a \im \dot z_t \bar z_t$, while the power extracted
by the two baths can be written as $\widetilde W^f_{t}= 2\re \left[
\dot z_t \left( \g_f \dot{\bar z}_t - \bar \r^f_{t}\right)\right]$ and
$\widetilde W^s_{t} = 2 \re \left[ \dot z_t \left( \frac{\g_s}{\t_s}
\bar z_t - \bar u_t \right) \right]$.

For concreteness we focus on the potential ${\cal
V}(|z|^2)=\frac{g}{2} |z|^4$ and compare with the results obtained for
the harmonic case, ${\cal V}(|z|^2)=k |z|^2$.  The simulation has been
performed for $\a=0.5$, $T_f=0.6$, $\g_f=1$, $T_s=2$, $\g_s=1$ and
$\t_s=1$; we set $k=1$ in the linear case and $g=1$ in the nonlinear
one.  The system (\ref{riscritte}) is numerically solved via a
standard discretization of the equations with time step $\d t =0.01$;
the noises are extracted using the routine {\sc gasdev} of the C
numerical recipes~\cite{C++}.

We found that ${\cal S}^V_\tau$ and ${\cal S}_\tau^{diss}$ are
uncorrelated (within the precision of the numerical data).  To the
extent that this is the case, their {\sc pdf}s can be studied
separately. Unfortunately, the {\sc pdf} of ${\cal S}^V_\tau$ is too
noisy to allow for a verification of the fluctuation relation in the
nonlinear case.  This is probably due to the fact that while in the
linear case ${\cal S}^V_\tau$ reduces to a boundary
term~\cite{newcomment}, in the non-linear case `spurious' boundary
contributions might be difficult to eliminate~\cite{VzC04}.  Indeed,
for the accessible values of $\t$, the variance of ${\cal S}^V_\tau$
is much larger than its average (while the fluctuation relation
predicts a variance of the order of $\s^V_+$). The large variance
might be a finite-$\t$ effect due to a boundary term with fluctuations
contributing to the ones of ${\cal S}^V_\tau$ but not to the
average. If this were the case, the fluctation relation should hold
for $|p|<1$ and very large $\t$. However, the required values of $\t$
might be so large to render the fluctuation relation unobservable in
practice, see~\cite{ZRA04a,BGGZ05}. For this reason, we shall not
discuss the data for ${\cal S}^V_\tau$.  The validity of the FR for ${\cal
S}^V_\tau$ (possibly minus a boundary term) in the nonlinear case
remains an open question that should be addressed by future work.

\subsection{Entropy production rate}

Let us now discuss
the behavior of ${\cal S}_\tau^{diss}$. The dissipative contribution to 
the entropy production rate, see Eq.~(\ref{eq:generic-result}), 
is given by
\beq 
\label{sigma3OK} 
\sigma^{diss}_t 
= 
\int_{-\io}^t dt' \  
T^{-1}(t-t') \, 2\a \im \big[ \dot z_t \bar z_{t'} +  
\dot z_{t'} \bar z_t \big] 
\ .
\eeq  
The inverse of the frequency dependent temperature, $1/T(\o)$, is 
\beq 
\label{TeffV}
\frac{1}{T(\o)} =  
\frac{\g_f (1 + \o^2 \t_s^2) 
+ 
\g_s}{T_f \g_f  (1 + \o^2 \t_s^2) + T_s \g_s} 
\; ,
\eeq 
see Eq.~(\ref{TeffNbaths}). 
Thus 
\beq 
\label{Tstar} 
T^{-1}(t) = \frac{1}{T_f} \delta(t) + \frac{\g_s}{T_f
\g_f\t_s^2}\left(1 - \frac{T_s}{T_f}\right) \frac{e^{-\Omega |t|}}{2
\Omega} \ , \;\;\;\;\mbox{with} \;\;\;\; \Omega =\frac{1}{\t_s}
\sqrt{\frac{T_f \g_f + T_s \g_s}{T_f \g_f}} 
\ ,  
\eeq 
and $T^{-1}(t)$
decays exponentially for large $t$.  Note that, if the bath is in
equilibrium, $T_s=T_f=T$, one has $T^{-1}(t)=\d(t)/T$ and
$\sigma^{diss}_t= 2 \a \im \dot z_t \bar z_t / T = W_t/T$ as expected
[recall that in our convention $\int_{-\io}^t dt' \, \d(t-t') =
\frac{1}{2} $].

 
\begin{figure} 
\centering 
\includegraphics[width=.75\textwidth,angle=0]{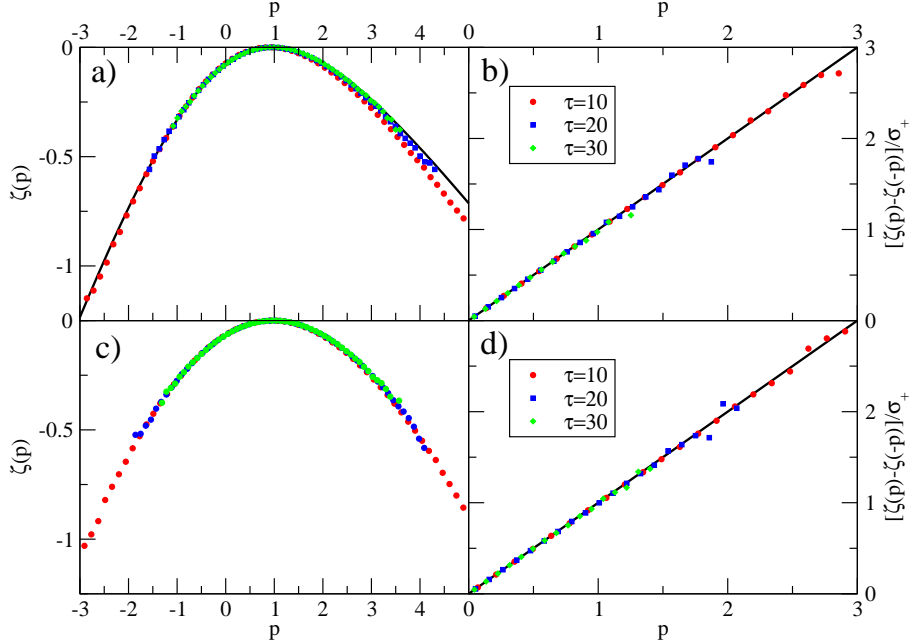} 
\caption{{\sc pdf} of $\sigma^{diss}_t$: a) The large deviation 
function for the 
harmonic potential at $\T=10,20,30$; the full line is the analytical 
result. 
b) The function $f(p) \equiv [\z_{diss}(p)-\z_{diss}(-p)]/\sigma^{diss}_+$ 
for the 
harmonic potential: the fluctuation theorem predicts a straight line 
with slope 1, represented by a full line.  c) The large deviation 
function for the quartic potential at $\T=10,20,30$.  d) The function 
$f(p)$ for the quartic potential: also in this case
the fluctuation theorem is well verified.} 
\label{fig:1} 
\end{figure} 
 
 
The data for ${\cal S}^{diss}_\tau$ are shown in Fig.~\ref{fig:1}.
The large deviation function $\z_{diss}(p)$ is reported in panel a) for the
harmonic and in panel c) for the quartic potential. The average
$\sigma^{diss}_+$ is equal to $0.332$ in the harmonic case and to
$0.276$ in the quartic case. The function $\z_{diss}(p)$ converges fast
to its asymptotic limit $\T \rightarrow \io$ (note that even the data
for $\tau\sim 10$ are in quite good agreement with the analytic
prediction for the harmonic case).  The fluctuation theorem predicts
$f(p) \equiv [\z_{diss}(p)-\z_{diss}(-p)]/\sigma^{diss}_+ = p$.  The function $f(p)$
is reported in panel b) for the harmonic and in panel d) for the
quartic potential.  In the harmonic case the numerical data are
compatible with the validity of the fluctuation theorem, as predicted
analytically.  Remarkably, the same happens in the quartic case for which 
we do not have an analytical prediction.

These results support the conjecture that, if ${\cal S}^{diss}_\tau$
and ${\cal S}^V_\tau$ are uncorrelated, the {\sc pdf} of ${\cal
S}_\tau^{diss}$ verifies the fluctuation theorem independently of the
form of the potential $V(x,y)$.

\subsection{Approximate entropy production rate}
\label{sec:approximate}

We also investigated numerically the fluctuations of the entropy
production ${\cal S}_\t^\Theta$ used in some numerical and experimental
studies~\cite{GGK01,CGHLP03,FM04,Ga04} and that we discussed in
section~\ref{sec:non-eq}. For this model it is given by 
\beq
\su_t=\frac{W_t}{\Th}=\frac{2\a}{\Th}\im \dot z_t \bar z_t 
\ .  
\eeq
Rather arbitrarily we set $\Th=T_f$ in the definition of $\su_t$.
This reflects what is usually done in numerical simulations, where the
dissipated power is divided by the ``kinetic'' temperature, {\it i.e.}
the temperature of the fast degrees of freedom.  Note that the choice
$\Th=T_f$ does not affect the function $\z_\Th(p)$ since the variable
$p$ is normalized [{\it i.e.}, $\z_\Th(p)\equiv \z(p)$ does not depend
on $\Th$, see Eq.~(\ref{pdef})] but it changes the average $\su_+$
that is proportional to $\Th^{-1}$.

 
\begin{figure} 
\centering 
\includegraphics[width=.75\textwidth,angle=0]{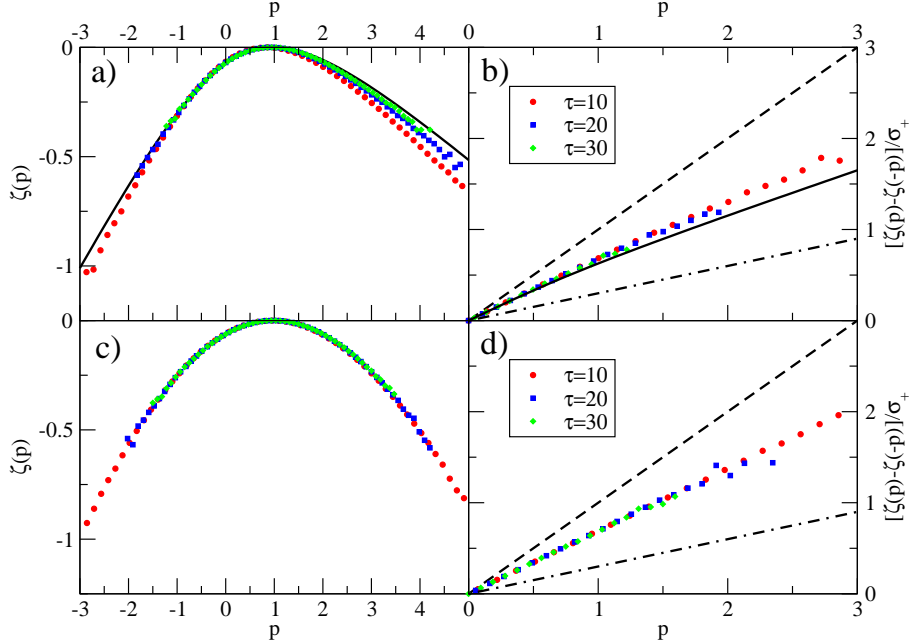} 
\caption{{\sc pdf} of $\su_t$: a) The large deviation function for the
harmonic potential at $\T=10,20,30$; the full line is the analytical
result.
b) The function $f(p) \equiv [\z(p)-\z(-p)]/\s^{T_f}_+$ for the
harmonic potential. The full line is the analytical prediction, the dashed
line is the fluctuation relation, the dot-dashed line has slope $T_f/T_s$.  
c) The large deviation function for the quartic potential at $\T=10,20,30$. 
d) The function $f(p)$ for the quartic potential; the dashed line
is the fluctuation theorem, the dot-dashed line has slope $T_f/T_s$.}
\label{fig:2} 
\end{figure} 
 
 
The data for $\su_t$ are reported in Fig.~\ref{fig:2}. The harmonic
case is shown in panels a) and b) while the anharmonic case is
presented in panels c) and d). We have $\s^{T_f}_+=0.455$ for the
harmonic potential and $\s^{T_f}_+=0.366$ for the quartic one. The
numerical result for the large deviation function of $\su_t$ agrees
very well with the analytical prediction in the harmonic case but, as
discussed in section~\ref{sec:non-eq}, it does not verify the
fluctuation theorem for $\Th=T_f$, as one can clearly see from the
right panels in Fig.~\ref{fig:2}.

Remarkably, in both the harmonic and anharmonic cases the function
$f(p) \equiv [\z(p)-\z(-p)]/\s^{T_f}_+$ is approximately linear in $p$
with a slope $X$ such that $1 > X > T_f/T_s$, {\it i.e.} $\z(p)-\z(-p)
\sim X \; p\s^{T_f}_+$.  If $f(p) \sim X p$, one can tune the value of
$\Th$ in order to obtain the fluctuation relation $\z(p)-\z(-p) = p
\su_+$, by simply choosing $\Th=\Th_{eff}=T_f/X$, thus defining a single
``effective temperature'' $\Th_{eff} \in [T_f,T_s]$.  From the data
reported in Fig.~\ref{fig:2} we get a slope $X \sim 0.66$, that gives
$\Th_{eff} = T_f/X \sim 0.9$.

\begin{figure} 
\centering 
\includegraphics[width=.55\textwidth,angle=0]{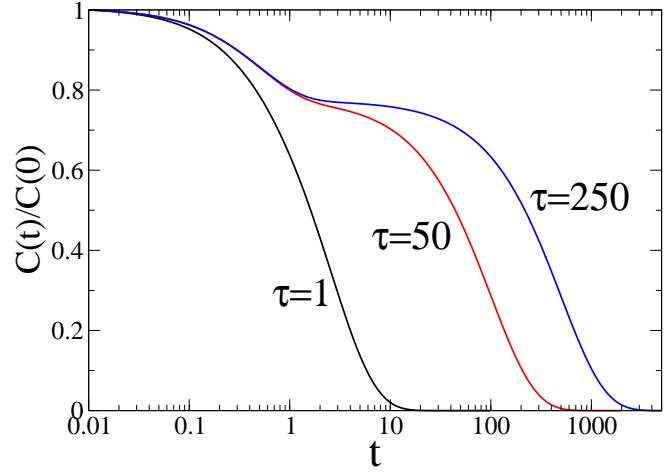} 
\caption{ 
Normalized autocorrelation functions of $z_t$ for 
the harmonic oscillator with 
$\a=0.5$, $k=1$, $T_f=0.6$, $T_s=2$, $\g_f=1$, $\g_s = k \t_s$ 
and $\t_s=1,50,250$. 
} 
\label{fig:3} 
\end{figure}

This behavior reflects the one found in some recent experiments
\cite{ZRA04b,GGK01,CGHLP03,FM04} in situations in which the dynamics
of the system happens essentially on a single time scale.  This is the
case also in our numerical simulation: in Fig.~\ref{fig:3} we report
the autocorrelation function $C(t) = \re \langle z_t \bar z_0 \rangle$
of $z_t$ (computed in Appendix~\ref{app:C}) for the harmonic
potential. The present simulation refers to the curve with $\t_s=1$,
which clearly decays on a single time scale.

\begin{figure} 
\centering 
\includegraphics[width=.55\textwidth,angle=0]{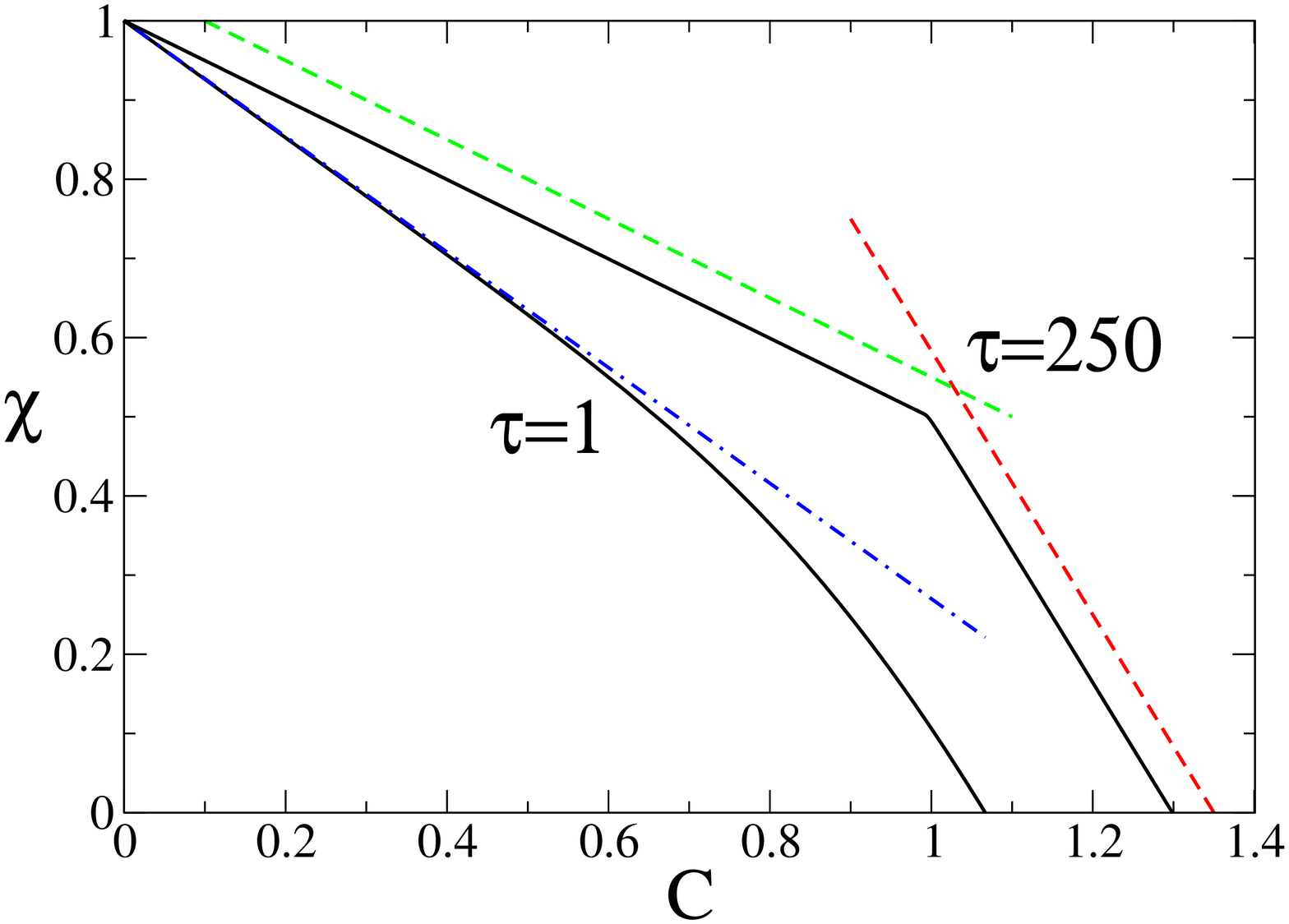} 
\caption{Parametric plot of the integrated response $\chi(t)$ as a
function of the correlation function $C(t)$ for the harmonic
oscillator with $\a=0$, $k=1$, $T_f=0.6$, $T_s=2$, $\g_f=1$, $\g_s = k
\t_s$ and $\t_s=1$ (continuous curve) and $\t_s=250$ (broken
curve). The dot-dashed line has slope $-1/1.37$, the dashed lines have
slope $-1/T_s$ and $-1/T_f$.}
\label{fig:3b} 
\end{figure}

In Fig.~\ref{fig:3b} we report the parametric plot $\chi(C)$ (see the
Introduction and Sect.~\ref{sec:II}) for the same set of parameters,
but $\a=0$.  The integrated response is given by $\chi(t)=\int_0^t dt'
\, R(t')$ and $R(t)$ is computed in Appendix~\ref{app:C}.  We see
that, for $\t_s=1$, the function $\chi(C)$ has a slope close to $-1/T_f$
at short times (corresponding to $\chi \sim 0$). For longer times, the
slope moves continuously toward $-1/T_{eff}$, with $T_{eff} \sim
1.37$.  This value of $T_{eff}$ is of the order of $\frac{\g_f T_f +
\g_s T_s}{\g_f+\g_s} =1.3$, which means that on time scales of the
order of the (unique) relaxation time the two baths behave like a
single bath equilibrated at an intermediate temperature. This would
be {\it exact} if the time scales of the two baths were exactly
equal.

It is worth to note that in this situation we get $T_{eff} \neq
\Th_{eff}$, that is, the effective temperature that one would extract
from the {\it approximate} fluctuation relation of Fig.~\ref{fig:2} does
not coincide with the effective temperature obtained from the
$\chi(C)$ plot of Fig.~\ref{fig:3b}. In particular, we get $T_f <
\Th_{eff} < T_{eff}$: this relation is consistent with the results of
\cite{ZRA04b} obtained from the numerical simulation of a sheared
Lennard-Jones--like mixture, even if the coincidence might be
accidental.


\subsection{Discussion}

Let us summarize the results in this section.
The numerical simulation of the non-linear problem confirms that the
fluctuation theorem is satisfied {\it exactly} when the entropy
production rate $\sigma_t^{diss}$ is defined using the power injected by
the external drive and the temperature in (\ref{eq:Tomega-def})
is used.

In situations in which the dynamics of the system happens on a single
time scale, a fitting parameter $\Th$ can be
introduced to obtain an {\it approximate} fluctuation relation
on the entropy production rate $W_t/\Th$.  However,
$\Th$ is not necessarily related to the effective temperature
$T_{eff}$ that enters the modified fluctuation--dissipation relation.
In the systems considered so far \cite{ZRA04b} one finds
$\Th<T_{eff}$. However, this is just an approximation that fails 
in more generic non-equilibrium situations.
In the next section we show that, when the dynamics happens on
different, well separated, time scales, it is impossible to find a
single value of $\Th$ such that $\su_t=W_t/\Th$ verifies the
fluctuation relation.

\section{Separation of time scales and driven glassy systems} 
\label{sec:VI} 

In this Section we discuss the application of our results to 
glassy systems. After presenting the general argument, we 
show explicitly that the dissipative entropy production 
satisfies the fluctuation relation for the  $p$-spin spherical 
model. We finally discuss an adiabatic approximation that 
allows one to derive approximate results in the case of 
systems with well-separated time-scales. 

\subsection{Background}
 
As discussed in the Introduction, in the study of mean-field models
for glassy dynamics~\cite{Cu02,Cukuja} and when using resummation
techniques within a perturbative approach to microscopic glassy models
with no disorder, effective equations of motion of the form of
Eq.~(\ref{motionglassy}) are obtained: 
\beq 
\label{motionglassy-rep}
\g \dot \f_t = -\mu(t) \ \f_t + \int_{-\io}^t dt' \ \Si(t,t') \
\f_{t'} + \r_t + \a h[\f_t] 
\ , 
\eeq 
where $\r_t$ is a Gaussian noise
such that $\langle \r_t \r_{t'} \rangle = D(t,t')$.  The
``self-energy'' $\Si(t,t')$ and the ``vertex'' $D(t,t')$ depend on the
interactions in the system throught the correlation and response of the 
field $\varphi$.  In absence of drive ($\a=0$) this
equation has a dynamic transition at $T_d$ separating a high temperature 
phase in which the dynamics rapidly equilibrates from a low temperature 
phase in which the dynamics 
is non--stationary and the fluctuation--dissipation relation is
violated. This means that the time needed for the system to equilibrate 
is the longest time scale that is unreachable in the calculation
(it already diverged with $N$). 

In the case of a driven mean field
system~\cite{Cukulepe,Cukulepe1,BBK00}, the external force is also
present in Eq.~(\ref{motionglassy-rep}) and after a transient the
system becomes stationary {\it for any temperature}, \ie $\mu(t)
\equiv \mu$, $\Si(t,t')=\Si(t-t')$, and $D(t,t')=D(t-t')$. The
functions $D$ and $\Si$ depend on the strength $\a$ of the driving
force.  In this case, Eq.~(\ref{motionglassy-rep}) resembles
Eqs.~(\ref{motioncomplex}) and (\ref{motion2baths}), and our results
of Sect.~\ref{sec:II} apply.  By analogy with
Eq.~(\ref{eq:Tomega-def}), the {\it effective temperature} is
defined in terms of $\Si$ and $D$, see below.
 
As discussed in \cite{Cukulepe,Cukulepe1,BBK00,BB00,BB02}, for small
$\a$ the system shows a completely different behavior above and below
$T_d$, reflecting  the presence of a dynamical transition at
$\a=0$.  Above $T_d$, the fluctuation-dissipation theorem holds 
in the limit of $\a \rightarrow 0$;
the transport coefficient related to the driving force $\a$ approaches
a constant value for $\a \rightarrow 0$ (the linear response holds
close to equilibrium) and the systems behaves like a ``Newtonian
fluid''.  Below $T_d$, the fluctuation-dissipation relation is
violated also in the limit $\a \rightarrow 0$ and the transport
coefficient diverges in this limit: the system is strongly
nonlinear. For a wide class of systems, see the Introduction and
Ref.~\cite{Cu02}, the relation between $D$ and $\Si$ takes a very
simple form in the limit $\a \rightarrow 0$: the effective
temperature $T_{eff}(t-t')$ defined from the ratio between induced integrated 
response and correlation (see the Introduction) is given by the
temperature of the bath $T$ for small $| \, t-t' \, |$ and by a constant 
$T_{eff} > T$ for large $| \, t-t' \, |$.
Thus, we expect that above $T_d$
(and for $\a \sim 0$) the system behaves as if coupled to a single
equilibrium bath (and the fluctuation theorem holds for the entropy
production rate defined as $\s_t = \a h[\f_t] \dot \f_t/T$), while
below $T_d$ the system behaves as if coupled to two baths acting on
different time scales and equilibrated at different
temperatures. 

As already remarked in the introduction, these single-spin 
equations of motion are valid for times $\t$ 
that are finite with respect to the
 size $N$. For these times the system behaves like independent spins moving
in a harmonic potential in contact with a nonequilibrium environment.
Thus, for ${\cal{S}}_\t$ with $\t$ in this regime 
the results of section~\ref{sec:II}  apply
and the correct definition of the entropy production rate is
given by Eq.~(\ref{sigma3}), \ie by the power injected by the external
force alone, divided by the frequency-dependent effective temperature.

\subsection{The spherical $p$-spin}

The (modified) fluctuation relation can be checked explicitly for the 
$p$-spin spherical model. This model realizes explicitly 
the behaviour described above. The asymptotic dynamics 
in the low temperature phase occurs in a region of phase space that is 
called the {\it threshold} and it is far from the equilibrium 
states~\cite{Cuku93} since times that grow with $N$ are needed to reach them. 
 
The effective equations of motion for the driven spherical 
$p$-spin~\cite{Cukulepe,BBK00} are 
\beq
\label{eq6:MCT}
\begin{split}
&\dpr_t C(t,t')=
-\m(t) C(t,t') + \int dt'' \Si(t,t'') C(t'',t') + 
\int dt'' D(t,t'') R(t',t'') \ , \\
&\dpr_t R(t,t')=-\m(t) R(t,t') + 
\int dt'' \Si(t,t'') R(t'',t') + \d(t-t') \ , \\
&\m(t)= T + \int dt' \big[ D(t,t') R(t,t') + \Si(t,t') C(t,t') \big] \ , \\
\end{split}
\eeq
with the vertex and self-energy
\beq \label{eq6:kernelrivisto}
\begin{split}
&D = \frac{p}{2} C^{p-1} + 
\a^2 \frac{k}{2} C^{k-1} \equiv D_0 + \a^2 D_1  \ ,\\
&\Si = \frac{1}{2}p(p-1)R C^{p-2} = R D'_0(C) \ ,
\end{split} 
\eeq
respectively. 
The dynamics can also be described with a
``single-spin'' Langevin equation of the form
\beq\label{eq6:MCTeffective}
\begin{split}
&\dot \varphi(t) = -\m(t) \varphi(t) +  
\int dt' \; \Si(t,t') \varphi(t') + \r(t) \ , \\
&\la \r(t) \r(t') \ra = 2T\d(t-t') + D(t,t') \ .
\end{split}\eeq
Note that $\Si$ and $D_0$ verify the detailed balance condition.
From the expressions (\ref{eq6:kernelrivisto}), one can rewrite 
Eq.~(\ref{eq6:MCTeffective}), in the stationary case, in the following way
\beq 
\label{motionglassy-rep2}
\begin{split}
&\dot \varphi(t) = 
-\mu \varphi(t) + 
\int_{-\io}^\io dt' \; \Si(t-t') \varphi(t') + \r(t) + 
\a h(t) \ , \\
&\la \r(t)\r(t') \ra = 2T\d(t-t') + D_0(t-t') \ , \\
&\la h(t) h(t') \ra = D_1(t-t') \ ,
\end{split}
\eeq 
where $\r(t)$ and $h(t)$ are two uncorrelated Gaussian variables.
Note that $\Si$ and $D_0$ still depend implicitly on $\a$ as one
has to solve the self-consistency equations (\ref{eq6:MCT}) for $C$ 
and $R$ and substitute the result in $\Si$ and $D_0$.

In the absence of drive and interpreting $\Sigma$ and $D$ as the
response and correlation of a bath, its 
frequency-dependent effective temperature is  
\beq
\label{Teff-pppp}
T(\o) = \frac{\n(\o)}{2\re g(\o)}= \frac{2T +
D_0(\o)}{2\re[1+\Si(\o)/(i\o)]} 
\ .  
\eeq 
Given that $\Sigma$ and $D$ depend on $R$ and $C$, if one finds that 
$R$ and $C$ are related
by the fluctuation dissipation theorem, 
$R(t) = -\b \th(t) \dot C(t)$, 
from the relation $\Si = R
D_0'(C)$ it follows that 
$T(\o) \equiv T$ and the bath is in equilibrium, as expected. If $R$
and $C$ do not verify the fluctuation dissipation theorem, $T(\omega)
\neq T$.

Switching on the external drive one can compute 
its dissipated power
\begin{equation}
W(t) = \a h(t) \dot \varphi(t)
\;  
\end{equation}
and its average. One finds 
\begin{eqnarray}
\label{Wpspin}
\la W(t) \ra &=& \a \la h(t) \dot \varphi(t) \ra = 
\a \la h(t) \Iint dt' \, \dot R(t-t') \big[ \r(t') + \a h(t') \big] \ra 
\nonumber\\
&=&
\la W \ra = \a^2 \int_0^{\io} dt \, \dot R(t) D_1(t) 
= \a^2 \frac{k}{2} \int_0^{\io} dt \, \dot R(t) C^{k-1}(t)
\ ,
\end{eqnarray}
consistently with the result of \cite{BBK00} where the average of the
injected power was explicitly computed for this model.  One can prove
that the entropy production ${\cal S}^{diss}_\t$ obtained from the rate 
\beq
\label{EPRhrandom}
\sigma_t^{diss}
= 
\a \int_{-\io}^t dt' \, T^{-1}(t-t') \big[ h(t) \dot \varphi(t')
+ h(t')\dot \varphi(t) \big] 
\ , 
\eeq 
verifies the fluctuation relation. Indeed, first rewriting 
the dissipative entropy production as 
\begin{equation}
{\cal S}^{diss}_\tau = \a
\frac{\D \o}{2 \pi} \sum_{n=-\io}^{\io} \frac{h_n i \o_n \bar
\varphi_n}{T(\o_n)} \ , 
\eeq
and using the solution for $\varphi_n$, in the generic
notation of the previous section,  one finds 
\beq
\phi_{diss}(\l) 
= 
\frac{1}{2} \Iint \frac{d\o}{2\p} \ln  \left[ 1 + 
4 \a^2 \l (1-\l) \frac{\m(\o) \o^2 [\re g(\o)]^2 }{\n(\o) |D(\o)|^2}
\right] \ ,
\eeq
where $\nu$, $\mu$ and $g$ are the Fourier transforms of 
the $\rho-\rho$ correlator, the $h-h$ correlator and the 
time-integrated $\Sigma$, respectively. 
It is easy to check that $\phi_{diss}(\l) = \phi_{diss}(1-\l)$
and the fluctuation relation is then verified. 

Once again, we showed that the dissipative contribution to the entropy 
production satisfies the fluctuation relation with a modified temperature. Note
however that this solvable example is non-trivial for at least two reasons.
It is a clearly an-harmonic problem since $\Sigma$ and $D$ are themselves 
functions of $C$ and $R$. The single spin $\varphi$ 
is coupled to an equilibrated bath at temperature
$T$ {\it and} a self-generated ``bath'' at a different temperature.
The temperature entering the fluctuation relation involves both, 
through the definition (\ref{Teff-pppp}). This temperature is 
the one that one would observe by measuring the  fluctuation-dissipation 
relation on the variable $\varphi$.


\subsection{The adiabatic approximation} 
\label{sec:adiabatic}
 
\begin{figure} 
\centering 
\includegraphics[width=.55\textwidth,angle=0]{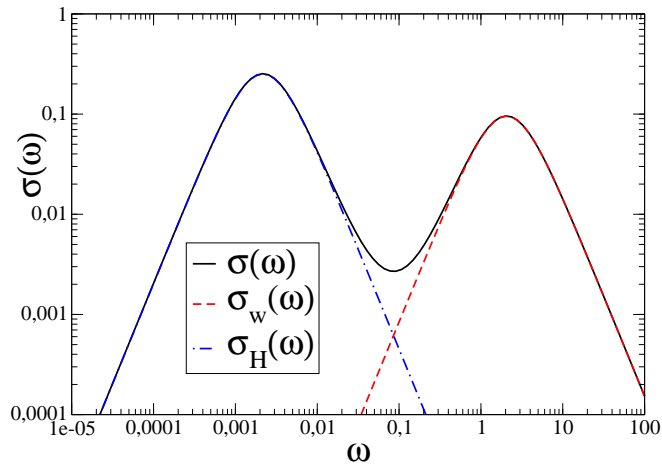} 
\caption{Power spectrum of the entropy production rate (full line) as
a function of the frequency for the harmonic oscillator with $\a=0.5$,
$k=1$, $T_f=0.6$, $T_s=2$, $\g_f=1$, $\g_s = k \t_s$ and $\t_s=250$.
The dot-dashed line is the ``slow'' contribution of $H_t$, the dashed
line is the ``fast'' contribution of $w_t$.  }
\label{fig:2b} 
\end{figure} 

When a simple system is coupled to a complex bath with two (or more)
time scales these are induced into the dynamics of the system.
When the time-scales are well separated, an adiabatic treatment 
is possible in which one separates the dynamic variables in terms 
that evolve in different time-scales (dictated by the baths) and
are otherwise approximately constant. 

In this subsection we use an adiabatic approach~\cite{Cukuja} to treat
simple problems coupled to baths that evolve on different scales.  The
motivation for studying this type of problems is that they resemble 
glassy systems although in the latter the separation of  
time-scales is self-generated.

We study the {\sc pdf} of ${\cal S}^{diss}_\tau$ and ${\cal S}_\tau$.
The former satisfies the fluctuation theorem (at least in the harmonic
case since corrections due to ${\cal S}^V_\tau$ might appear in
non-harmonic problems). We check that the adiabatic approximation does
not spoil this feature. The latter, instead, does not satisfy the
fluctuation theorem in general.  However, it is interesting to
understand under which conditions it satisfies the fluctuation theorem
approximately.  Indeed, in numerical simulations and experiments it
has been customary to measure the dissipated power $W_t = \a h[\f_t]
\dot \f_t$ and then define the entropy production rate as $\s_t=W_t/T$
(where $T$ is the temperature of the fast bath). This corresponds to
the definition of $\su_t$, given in Eq.~(\ref{sigma1}) for the
harmonic oscillator. We show that when the particle is coupled to a
bath that evolves in well-separated time-scales $\su_t$ {\it does not}
satisfy the fluctuation theorem.

\subsubsection{The harmonic model coupled to two baths}

Let us consider again the Langevin equation (\ref{motion2baths}) with
${\cal V}(|z|^2)=k |z|^2$. In this case, the correlation functions can
be calculated explicitly, see Appendix~\ref{app:C}. In
Fig.~\ref{fig:3} we report the autocorrelation functions,
$C(t)=\langle z_t \bar z_0 \rangle$, for $\a=0.5$, $k=1$, $T_f=0.6$,
$T_s=2$, $\g_f=1$, $\g_s = k \t_s$ and different values of $\t_s$.
Clearly, for $k \t_s = \g_s \gg \g_f$ two very different time scales
--related to the time scales of the two baths-- are present.  From the
plot of Fig.~\ref{fig:3b} one sees that in the case $k \t_s = 250 \gg
\g_f$ the function $\chi(C)$ is a broken line with slope $-1/T_f$ at
large $C$ (short times) and $-1/T_s$ for small $C$ (large times).

We want to show that, in this situation, the variable $z_t$ can be 
written as the sum of two quasi-independent contributions.
Using the construction introduced
in~\cite{Cukuja} we rewrite the equation of motion~(\ref{motion2baths})
as 
\begin{eqnarray}
\left\{ 
\begin{array}{l} 
\g_f \dot z_t = - (k + \g_s/\t_s) z_t + i\a z_t + \r^f_{t} + h_t 
\; , 
\\ 
h_t= - \frac{\g_s}{(\t_s)^2}  
\int_{-\io}^t dt' \ e^{-\frac{t-t'}{\t_s}} z_{t'} + \r^s_{t} 
\; .
\label{eq:second} 
\end{array}
\right. 
\end{eqnarray}
The variable $h_t$ is ``slow''; if we consider it as a constant in the
first equation, the variable $z_t$ will fluctuate around the
equilibrium position $z_h = h / (k + \g_s/\t_s - i \a) \equiv H$. The
latter will --slowly-- evolve according to the second equation
in~(\ref{eq:second}), in which we can approximate $z_{t'} \sim
H_{t'}$. Defining the --fast-- displacement of $z_t$ with respect to
$H_t$, $w_t \equiv z_t - H_t$, we obtain the following equations for
$(w_t,H_t)$: 
\beq
\label{adiabatic2} 
\begin{cases} 
&\g_f \dot w_t = - (k + \g_s/\t_s) w_t + i\a w_t + \r^f_{t} \ , \\ 
&\frac{\g_s}{\t_s}
\int_{-\io}^t dt' \ 
e^{-\frac{t-t'}{\t_s}} \dot H_{t'} = -k H_t +i\a H_t + \r^s_{t} \ . 
\end{cases} 
\eeq 
In this approximation, $z_t = H_t + w_t$ is the sum of two
contributions: $w_t$ is a ``fast'' variable which evolves according to
a Langevin equation with the fast bath only and a renormalized
harmonic constant $k+\g_s/\t_s$, while $H_t$ is a ``slow'' variable
which evolves according to a Langevin equation where the slow bath
only appears. In both equations the driving force $\a$ is present,
thus we expect both $H_t$ and $w_t$ to contribute to the dissipation.
Note that $w_t$ and $H_t$ are completely uncorrelated in this
approximation.
 
\subsubsection{The ``potential'' entropy production rate 
$\s^V_t$}

In the adiabatic approximation the term ${\cal{S}}^V_\t$ in
equation~(\ref{eq:generic-result}) should become a boundary
term. Indeed, the function $T^{-1}(t)$, in the adiabatic
approximation, becomes 
\beq 
T^{-1}(t)=\frac{1}{T_f}\d(t) + T^{-1}_s(t)
\ , 
\eeq 
where the function $T_s^{-1}(t)$ is ``slow'', see {\it e.g.}
Eq.~(\ref{Tstar}).  Inserting this expression in $\s^V_t$, the first
term gives a total derivative.  The second term gives 
\beq
\int_{-\io}^t dt' \, T^{-1}_s(t-t') \left[ \dot r_\a(t) 
\frac{\partial V}{\partial r_\a}[\vec r(t')] + 
\dot r_\a(t') \frac{\partial V}{\partial r_\a}[\vec r(t)] \right] 
\ .  
\eeq 
Due to the convolution with the ``slow''
function $T^{-1}_s(t)$, the fast components of $r$ are irrelevant in
the integral, while for the slow ones it is reasonable to replace
$\dot r_\a(t)$ with $\dot r_\a(t')$ on the scale $\t_s$ over which
$T^{-1}_s(t)$ decays.  Thus one obtains a total derivative times
the integral of $T^{-1}_s(t)$ which is a finite constant.  Obviously
this is not a rigorous proof and should be checked numerically in
concrete cases.

\subsubsection{The ``dissipative'' entropy production rate $\sigma^{diss}_t$} 

The entropy production rate defined in
Eqs.~(\ref{sigma3}) and~(\ref{sigma3OK}) can be rewritten in terms of
$H_t$ and $w_t$. Recalling that $T^{-1}(t)$ is defined by Eq.~(\ref{Tstar})
one obtains (the details of the calculation are reported in
Appendix~\ref{app:D}) 
\beq
\label{sigma3adiabatic} 
\sigma^{diss}_t \sim 2\a \im \left[ \frac{\dot w_t \bar w_t}{T_f} + \frac{\dot
H_t \bar H_t}{T_s} \right] 
\eeq 
neglecting terms that vanish when
$\sigma^{diss}_t$ is integrated over time intervals of the order of $\t_s$. This
is exactly the entropy production expected for two independent
systems.  

\begin{figure} 
\centering 
\includegraphics[width=.55\textwidth,angle=0]{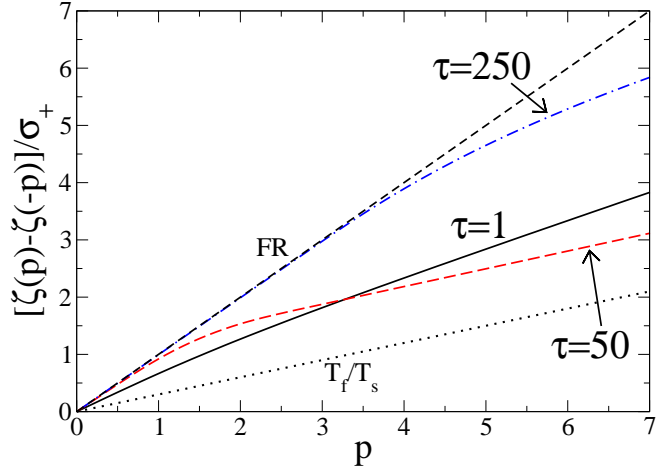} 
\caption{The function $[\z_\Th(p) - \z_\Th(-p) ] / \su_+$ for the
harmonic oscillator with $\a=0.5$, $k=1$, $\Th=T_f=0.6$, $T_s=2$,
$\g_f=1$, $\g_s = k \t_s$ and $\t_s=1,50,250$. The dashed line (FR) is a
straight line with slope 1, the dotted line has slope $T_f/T_s=0.3$.
}
\label{fig:4} 
\end{figure}

To check that this approximation works well, let us introduce a 
``power spectrum'' $\s(\o) d\o$ as the contribution coming from frequencies
$[\o,\o+d\o]$ to the average entropy production rate, 
$\sigma^{diss}_+ = \int_0^\io d\o \, \s(\o)$.
From Eq.~(\ref{phi_eff}) we get
\beq
\label{powersp}
\begin{split}
&\sigma^{diss}_+ = \left.\frac{d \phi_{diss}}{d\l}\right|_{\l=0} = 
-\int_{-\io}^{\io} \frac{d\o}{2\p} \frac{ 2 \a\o \n(\o)}{|D(\o)|^2} \\
&\s(\o)=\frac{1}{2\pi} \left[ - \frac{ 2 \a\o \n(\o)}{|D(\o)|^2} + \frac{ 2 \a\o \n(-\o)}{|D(-\o)|^2}
\right] = \frac{\a\o \n(\o)}{\pi} \left[ \frac{1}{|D(-\o)|^2} - \frac{1}{|D(\o)|^2}\right]
\end{split}
\eeq 
Substituting the expressions of $\n(\o)$ and of $D(\o)$
appropriate for Eq.~(\ref{motion2baths}) we get the power spectrum
$\s(\o)$ as a function of $\o$ which is reported in Fig.~\ref{fig:2b}
as a full line.  The contributions of $w_t$ and $H_t$, $\s_w(\o)$ and
$\s_H(\o)$, are obtained inserting in Eq.~(\ref{powersp}) the
expression of $\n(\o)$ and $D(\o)$ obtained from the two equations
(\ref{adiabatic2}). They are reported as dashed and dot-dashed lines
in Fig.~\ref{fig:2b}.  We can conclude that, for $k \t_s \gg \g_f$,
the adiabatic approximation holds and $\sigma^{diss}_t = \s^w_t +
\s^H_t$, with $\s^w_t = 2\a \im \, \dot w_t \bar w_t / T_f$ and
$\s^H_t = 2\a \im \, \dot H_t \bar H_t / T_s$, and the two
contributions are independent.  Note that the average dissipation due
to $H$ is much larger than the one due to $w$.  Finally, we can write:
\beq 
\phi_{diss}(\lambda) = \phi^w(\lambda) + \phi^H(\lambda) \ ,
\hspace{30pt} \phi^{w,H}(\l) = -\lim_{\t \rightarrow \infty} \t^{-1} \ln
\langle \exp\big[-\l \s^{w,H}\big] \rangle \ .  
\eeq 
Both $\phi^w(\lambda)$
and $\phi^H(\lambda)$ verify the fluctuation theorem, as the two equations of
motion (\ref{adiabatic2}) are particular instances of the general case
discussed in section~\ref{sec:II}. The function $\z_{diss}(p)$ is the
Legendre transform of $\phi_{diss}(\l)$ and will verify the fluctuation
theorem.

\subsubsection{The ``approximate'' entropy production rate $\su_t$}
  
In the same approximation, $\su_t$ is given, for $\Th=T_f$, by 
\beq 
\label{sigma1adiabatic} 
\s^{T_f}_t = 2\a \im \left[ \frac{\dot w_t \bar w_t + 
\dot H_t \bar H_t}{T_f} \right] 
= \s^w_t + \frac{T_s}{T_f}\s^H_t  \ , 
\hspace{20pt} \text{and} \hspace{20pt} \phi_{T_f}(\l) =
\phi^w(\lambda) + \phi^H(\lambda T_s / T_f) \ ; \eeq the contribution
of $H_t$ is weighted with the ``wrong'' temperature, \ie the
temperature of the fast degrees of freedom. Indeed, as we have already
discussed, $\phi_\Th(\lambda)$ does not verify the fluctuation
theorem.  The function $f(p) \equiv [\z_\Th(p) - \z_\Th(-p) ] / \su_+$
[obtained from Eq.~(\ref{phi_theta})] is reported in Fig.~\ref{fig:4}.
As already discussed in section~\ref{sec:V}, when the time scales of
the two baths are of the same order, $k \tau_s \sim \g_f$, the two baths act
like a single bath at temperature $\Th \in [T_f,T_s]$ and the function
$f(p)$ is approximately linear in $p$ with slope $X \in [T_f/T_s,1]$.
When the time scales are well separated, $k\t_s \gg \g_f$, the
adiabatic approximation holds; and one finds that $f(p)$ has
slope $\sim 1$ for small $p$ and $T_f/T_s$ for large $p$
(see Fig.~\ref{fig:4}).

The results for $k \t_s \sim \g_f$ are consistent with the ones in
\cite{ZRA04b} where only the situation in which the two time scales
are not well separated could be investigated. Indeed, when the
dynamics becomes very slow the observation of large negative
fluctuations of the entropy production requires a huge amount of
computational time and the function $f(p)$ can be calculated only in a
narrow range of $p$ around $p=0$. Note that the value of $p$ at which
the slope of $f(p)$ crosses over from $1$ to $T_f/T_s$ depend on the
values of the parameters 
$\a$, $\g_f$, $\g_s$, $\t_s$, {\it etc.} and can be of the order of
$5$, while in numerical simulations one can usually reach values of --
at most -- $p \sim 3$. Thus, the observation of curves like the one
reported in Fig.~\ref{fig:4} in numerical simulations of glassy
systems is a very difficult task.
 
\section{Green-Kubo relations} 
\label{sec:Green-Kubo} 
 
It was proven in \cite{Ga96a,Ga96b} that the fluctuation theorem implies, in
the equilibrium limit ($\s_+ \rightarrow 0$) the Green-Kubo relation for
transport coefficients. This is a particular form of the
fluctuation--dissipation theorem.  In this Section we discuss how one links
the modified fluctuation theorem -- in which we replaced the external bath
temperature by the (frequency dependent) 
effective temperature of the unperturbed system -- to the
modification of the fluctuation dissipation relation.
 
\subsection{General derivation}

Let us recall briefly how the Green-Kubo relation can be obtained from the
fluctuation theorem. Suppose that we apply a (constant) driving force $E$ to
a system in equilibrium. This generates a corresponding flux $J_t$ (\eg if $E$
is an electric field $J_t$ is the electric current) such that, close to
equilibrium, the dissipated power can be written as $W_t = E J_t$.  The
entropy production rate is then 
\beq
\label{sclosetoeq} 
\s_t = \frac{W_t}{T} = \frac{E J_t}{T} 
\ .  
\eeq 
The fluctuation
theorem can be written as $\phi(\l)=\phi(1-\l)$ where $\phi(\l)$ is
defined in Eq.~(\ref{phidef}). The derivatives of $\phi(\l)$ are the
moments of ${\cal S}_\tau$, \ie 
\beq 
\label{eq:moments} 
\phi_k \equiv
\left. \frac{d^k \phi}{d\l^k} \right|_{\l=0} = (-1)^{k-1}
\lim_{\t\rightarrow\io} \t^{-1} \langle {\cal S}_\t ^k \rangle_c \ ,
\eeq 
where $\langle A^2 \rangle_c = \langle A^2 \rangle - \langle A
\rangle^2$ and so on.  Thus, $\phi_k \sim \s_+^k$, and close to
equilibrium ($\s_+ \sim 0$) $\phi(\l)$ is well approximated by a
second order polynomial (corresponding to a Gaussian {\sc pdf}), 
\beq
\phi(\l) \sim \phi_0 + \phi_1 \l + \frac{\phi_2}{2} \l^2 
\ .  
\eeq
Then the fluctuation relation, $\phi(\l)=\phi(1-\l)$, implies $\phi_2
= -2\phi_1$; from Eq.~(\ref{eq:moments}) and 
using time-translation invariance,
\beq 
\phi_2 = -2 \phi_1
\hspace{10pt} \Rightarrow \hspace{10pt} 
\s_+ = \int_0^\io dt \ \langle \s_t \s_0 \rangle_c \ . 
\eeq  
Substituting $\s_t=E J_t/T$ one obtains 
\beq 
\langle J \rangle_E = 
\frac{E}{T} \int_0^\io dt \  \langle J_t J_0 \rangle_{E=0} + o(E^2) \ , 
\eeq 
that is to say,  the Green-Kubo relation. 

Note that even out of equilibrium one can define a flux 
${\cal J}_t$ using $\s_t$ as a ``Lagrangian'', see Ref.~\cite{Ga04}: 
\beq 
{\cal J}_t = \frac{\partial \s_t}{\partial E} \ . 
\eeq 
Close to equilibrium $\s_t$ is given by 
Eq.~(\ref{sclosetoeq}) and ${\cal J}_t = J_t/T$. 
If, in the absence of a drive, 
the system has a non trivial effective temperature, 
the entropy production rate should be defined as in Eqs.~(\ref{sigma3})
and (\ref{sigma3OK}).  
Then the flux ${\cal J}_t$ is given by 
\beq 
{\cal J}_t = \frac{\partial \sigma^{diss}_t}{\partial \a} =  
4 \, \im \int_{-\io}^t dt' T^{-1}(t-t') \dot z_t \bar z_{t'} = 
2 \int_{-\io}^t dt' T^{-1}(t-t') [\dot y_t x_{t'} - \dot x_t y_{t'}] \ . 
\eeq 
The fluctuation theorem for $\sigma^{diss}$ implies then a Green-Kubo 
relation for ${\cal J}_t$: 
\beq 
\label{GKgen} 
\langle {\cal J} \rangle_\a = \a \int_0^\io dt \ \langle {\cal J}_t {\cal J}_0
\rangle_{\a=0} + o (\a^2) \ .  \eeq The physical meaning of the latter
relation becomes clear if one writes the flux ${\cal J}_t$ in the adiabatic
approximation discussed in the previous section; from
Eq.~(\ref{sigma3adiabatic})
\beq {\cal J}_t = 2 \im \left[ \frac{\dot w_t
\bar w_t}{T_f} + \frac{\dot H_t \bar H_t}{T_s} \right] = \frac{J^w_t}{T_f} +
\frac{J^H_t}{T_s} \ , \eeq 
and Eq.~(\ref{GKgen}) becomes 
\beq \frac{\langle
J^w \rangle_\a}{T_f} + \frac{\langle J^H \rangle_\a}{T_s} = \frac{\a}{T_f^2}
\int_0^\io dt \ \langle J^w_t J^w_0 \rangle_{\a=0} + \frac{\a}{T_s^2}
\int_0^\io dt \ \langle J^H_t J^H_0 \rangle_{\a=0} + o(\a^2) \ .  
\eeq 
Indeed,
in the adiabatic approximation the Green-Kubo relation holds separately for
$J^w_t$ (with temperature $T_f$) and for $J^H_t$ (with temperature $T_s$).
Eq.~(\ref{GKgen}) encodes the two contributions and holds even when the
adiabatic approximation does not apply and the contributions of the ``fast''
and of the ``slow'' modes is not well separated.
 
Note that the ``classical'' Green-Kubo relation involves the total flux 
$J_t = J^w_t + J^H_t$. For the latter one has, in the adiabatic approximation,
\beq 
\label{GKgenFDR} 
\begin{split} 
\langle J_t \rangle_\a &= 
\langle J^w_t \rangle_\a + \langle J^H_t \rangle_\a = 
\frac{\a}{T_f} \int_0^\io dt \  \langle J^w_t J^w_0 \rangle_{\a=0} +  
\frac{\a}{T_s} \int_0^\io dt \  \langle J^H_t J^H_0 \rangle_{\a=0} \\ 
&=\a \int_0^\io dt \left[ \frac{\langle J^w_t J^w_0 \rangle_{\a=0}}{T_f} +  
\frac{\langle J^H_t J^H_0 \rangle_{\a=0}}{T_s} \right] \sim 
\a \int_0^\io dt \frac{1}{T_{eff}(t)} \langle J_t J_0 \rangle_{\a=0} \ . 
\end{split} 
\eeq 
The latter relation is the generalization of the Green-Kubo
formula that comes from the generalized fluctuation-dissipation
relation discussed in the Introduction and Section~\ref{sec:II}. It is
closely related, but not equivalent, to Eq.~(\ref{GKgen}).
 
\subsection{The Green-Kubo relation for driven glassy systems} 
 
Eqs.~(\ref{GKgen}) and (\ref{GKgenFDR}) cannot be applied
straightforwardly to driven glassy systems as for these systems the correlation
function $\langle J_t J_0 \rangle_\a$ is not stationary at $\a=0$ at low
temperatures.  Indeed, the relaxation time of the latter grows very fast 
as $\a
\rightarrow 0$ and at some point falls outside the experimentally accessible
range: the system will not be able to reach stationarity on the experimental
time scales and will start to {\it age} indefinitely.
 
However, let us consider again the equation of 
motion~(\ref{motionglassy}) for $\a \neq 0$, where we assume that a 
stationary state is reached,  
\beq 
\label{staglass} 
\g \dot \f_t 
= 
-\mu_\a \ \f_t + \int_{-\io}^t dt' \ \Si_\a(t-t') \ \f_{t'} + \r_t + 
\a h[\f_t] \ , 
\hspace{20pt} 
\langle \r_t \r_{t'} \rangle = D_\a(t-t') \ . 
\eeq  
The functions $\Si_\a(t-t')$ and $D_\a(t-t')$ depend (strongly) on $\a$;
indeed, as the term explicitly proportional to $\a$ is a small perturbation
for $\a \sim 0$, the main contribution to the $\a$-dependence of the 
dynamics of $\f_t$
will come from the $\a$-dependence of $\Si_\a$ and $D_\a$. If we compare the
latter equation with Eq.~(\ref{motion2baths}), we see that setting $\a = 0$ in
Eq.~(\ref{motion2baths}) is equivalent to setting $\a=0$ {\it without}
changing the functions $\Si$ and $D$ in Eq.~(\ref{staglass}). This will not
affect too much the correlation function $\langle J_t J_0 \rangle_\a$ if $\a$
is small. Finally, we can write, for small $\a$, 
\beq \langle J_t \rangle_\a
\sim \a \int_0^\io dt \frac{1}{T_{eff}(t)} \langle J_t J_0 \rangle_{\a} \ ,
\eeq 
even if the limit $\a \rightarrow 0$ is not well defined. An analogous
relation will be obtained from Eq.~(\ref{GKgen}) (which is equivalent to the
fluctuation theorem in the Gaussian approximation) within the same
approximation.  The latter relations can be tested in numerical simulations as
well as in experiments.

\section{Slow periodic drive and effective temperature}
\label{sec:XI}

In this Section we discuss a means to measure an effective temperature 
associated to slow timescales
of a non-equilibrium system by using the modification of the 
fluctuation theorem. 

A lesson we learn from the previous calculations (see
{\it e.g.}~Fig.~\ref{fig:2b}) is that the work done at large frequencies is
overwhelmingly larger than that done at very low frequencies --
precisely the one we wish to observe in order to detect effective
temperatures. One way out of this is to choose a perturbation that
does little work at high frequencies: a periodically
time-dependent force that {\em derives from a potential} $\cos(\Omega
t) {\tilde V}(r)$, with $1/\Omega$ of the order of timescale of the
slow bath $\tau_s$.  Let us show this for a one dimensional system,
the generalization being straightforward.

\subsection{General derivation}

Let us consider a single degree of freedom $r$ moving in a time-independent
potential $V(r)$ and subject to a periodically time-dependent field 
$\cos(\Omega t) {\tilde V}(r)$, and in
contact with a `fast' and a `slow' bath with friction kernel, thermal
noise and temperature $(\rho^f,g_f,T_f)$ and $(\rho^s,g_s,T_s)$,
respectively:
\begin{eqnarray}
m\ddot r(t) + \int_{-\infty}^{t}  dt' \;  g_f(t-t') {\dot r}(t') 
 + \int_{-\infty}^{t}  dt' \;  g_s(t-t') {\dot r}(t') =
-\frac{\partial V}{\partial r}[r(t)] + \rho^f(t) + \rho^s(t) - \cos(\Omega t) 
\frac{\partial {\tilde V}}{\partial r}[r(t)]
\; ,
\label{uno1} 
\end{eqnarray}
The time scale of the time dependent field $1/\Omega$ is of the same
order as the one of the `slow' bath.  The work in an interval of time
$(0,\t)$ done by the time-dependent potential is:
\begin{equation}
W_\t=-\int_0^\t \; \cos(\Omega t') \;
\frac{\partial  {\tilde V}}{\partial r}  \;
\dot r  \; dt' = - {\tilde V}(\t)+{\tilde V}(0) +
\Omega \int_0^\t  \; \sin(\Omega t')  \; {\tilde {V}}  \; dt' \ .
\label{work}
\end{equation}
Only the last term grows with the number of cycles, so for long times we can
 neglect the first two.
Now, integrating (\ref{uno1})  by parts, we obtain
\begin{eqnarray}
m\ddot r(t)   &=& -\int_{-\infty}^{t}  dt' \;  g_f(t-t') {\dot r}(t') 
-\frac{\partial V}{\partial r}[r(t)]  + \rho^f + h(t) - \hat h(t)
\frac{\partial {\tilde V}}{\partial r}[r(t)]
\label{dos}
\\
h(t) &\equiv& - \int_{-\infty}^{t}  dt' \; g_s(t-t')  r(t')  + \rho^s(t)
\; .
\label{tres} 
\end{eqnarray}
where $\hat h(t) = \cos(\Omega t)$.  In the adiabatic limit when both
the timescales of the slow bath and the period $1/\Omega$ of the
potential $\tilde V$ are large, $ h(t)$ and $\hat h(t)$ are
quasi-static.  Hence, $r$ has a fast evolution given by
Eq.~(\ref{dos}) with $h$ and $\hat h$ fixed and it reaches a
distribution~\cite{Cukuja}
\begin{equation}
P(r/h,\hat h)= \frac{ 
e^{-\beta_f \left( V+\hat h {\tilde V} + g_f(0) \frac{r^2}{2} - hr \right)}
}
{
\int \; dr \; 
e^{-\beta_f \left( V+\hat h {\tilde V} +  g_f(0) \frac{r^2}{2} - hr \right) }
}
\; .
\label{ll}
\end{equation}
The denominator defines $Z(h,\hat h)$ and 
$F(h,\hat h) \equiv -\beta_f^{-1} \ln Z(h,\hat h)$. 
Note that $F(h,\hat h(t))$ is periodically time-dependent
 through $\hat h$.
The approximate evolution of $h$ 
is now given by Eq.~(\ref{tres}) with the replacement of 
$r$ in the friction term by its average $\frac{\partial F(h,\hat h)}{\partial h}$
with respect to the fast evolution:
\begin{equation}
h(t) = \int_{-\infty}^{t} dt' \, g_s(t-t')
\frac{\partial F(h,\hat h)}{\partial h}(t')  + \rho^s(t)
\; .
\label{cinco} 
\end{equation}
Eq. (\ref{cinco}) is in fact a generalized Langevin
 equation for a  system coupled to a (slow) bath
of temperature $T_s$. Indeed, it can be shown \cite{Cukuja} to be equivalent to 
a set of degrees of freedom $y_i$ evolving according to
the ordinary Langevin equation:
\begin{equation}
\left[ m_j \frac{d^2}{dt^2} + \gamma_j \frac{d}{dt} + \Omega_j \right] y_j = \xi_j(t)- 
\frac{\partial F \left( \sum_j A_j y_j \right) }{\partial y_j}
\label{truc}
\end{equation}
with $ \langle \xi_i(t) \xi_j(t') \rangle = 2 T_s \gamma_j \delta_{ij}
\delta(t-t') $, provided that the Fourier transforms $g_s(\omega)$ an
$\nu_s(\omega)$ of friction kernel and noise autocorrelation can be
written as:
\begin{equation}
g(\omega) =
\sum_j \frac{ A^2_j}{m_j (\omega-\omega_j^+)(\omega-\omega_j^-)} \qquad \qquad
\nu(\omega) = 
2 T_s \sum_j 
\frac{ \gamma_j A^2_j}{m^2_j (\omega-\omega_j^+)(\omega-\omega_j^-)
(\omega+\omega_j^+)(\omega+\omega_j^-)}
\end{equation}
where $\omega^\pm_j$ are the roots of $-m_j \omega^2+i\gamma_j\omega +
\Omega_j=0$.

Within the same approximation leading to (\ref{cinco}),  
 the average of $\tilde V(r)$ over a time window $\Delta$ that
 is long compared to the short timescale, but
sufficiently slow that we can consider that $h$ and $\hat h$ are constant is
\begin{equation}
\int_t^{t+\Delta}   \;  
\tilde V(r(t')) \; dt' \sim \Delta \int \; dr \;  P(r/h,\hat h) \; 
\tilde V(r) =
\Delta \frac{\partial F(h,\hat h)}{\partial \hat h}
\end{equation}
so that we obtain
\begin{equation}
W_\t \sim  \Omega \int_0^\t \sin(\Omega t')  
\frac{\partial F(h,\hat h(t'))}{\partial \hat h} dt' =
- \int_0^\t \; \frac{\partial F(h,\hat h(t'))}{\partial t'} dt'
\label{work1}
\end{equation} 
which tells us that for long time intervals the work done by the
original time-dependent potential $\tilde V$ is indeed the same as the
work done by the time-dependent effective potential $F$ in
(\ref{cinco}).

The fluctuation theorem then holds for the distribution of this work,
with a single temperature $T_s$.  We conclude that the distribution of
work due to a slow perturbation satisfies the fluctuation theorem with
only the slow temperature, and can be hence used experimentally to
detect it.

\subsection{Experimental realization}

The simplest application of the above general result is obtained
considering $\tilde V(r)=\tilde h r$ and $V(r)=k r^2$. Then, grouping
together the two noises in a single one with friction $g = g_f + g_s$
and correlation $\n = T_f \n_f + T_s \n_s$ as described in
Sect.~\ref{sec:II}, Eq.~(\ref{uno1}) simply becomes 
\beq 
m \ddot r(t)
+ \int_{-\io}^t dt' \, g(t-t') \dot r(t') = -k r(t) + \r(t) + \tilde h
\cos (\O t) \ .  
\eeq 
This equation describes for instance the motion
of a Brownian particle moving in an out of equilibrium environment and
trapped by an harmonic potential whose center oscillates at frequency
$\O$.  A concrete experimental realization of this setting has been
already considered in \cite{AG04}: Silica beads of $\sim 2 \m$m
diameter were dispersed in a solution of Laponite (a particular clay
of $\sim 30 $nm diameter) and water.  The Laponite suspension form a
glass for large enough concentration of clay and provides the
nonequilibrium environment. The Silica beads are Brownian particles
diffusing in such environment. They can be trapped by optical
tweezing, and the center of the trap can oscillate with respect to the
sample if the latter is oscillated through a piezoelectric stage. In
\cite{AG04} the mobility and diffusion of tracer particles were
measured obtaining an estimate of $T_{eff}(\O)$.  Here we propose to
measure the work done by the trap on the tracers.
Indeed, the work dissipated in $(0,\t)$ is linear in $r(t)$ so it
should be possible to measure it simply through the measurement of
$r(t)$: \beq
\label{workharmonic}
W_\t = \O \tilde h \int_0^\t dt' \sin (\O t') r(t') \ ;
\eeq
note that, as $W_\t$ is linear in $r(t)$, it is a Gaussian variable.
With a simple calculation one finds
\beq
\lim_{\t \rightarrow \io}
\frac{\langle (W_\t - \langle W_\t \rangle)^2 \rangle}{2\langle W_\t \rangle} =  
\frac{\n(\O)}{2 \re g(\O)} = T_{eff}(\O)
\eeq
This means that the (Gaussian) {\sc pdf} of ${\cal S}^{diss}_\t = W_\t / T_{eff}(\O)$ 
satisfies the fluctuation relation. If the two baths are modeled as in 
Sect.~\ref{sec:V}
with $k \g_s = \t_s \gg \g_f$ one has $T_{eff}(\O) = T_s$ for $\O \t_s < 1$ 
[see Eq.~(\ref{TeffV})].
The measurement of the distribution of the work (\ref{workharmonic}) allows for
the measurement of $T_s$. Note that other experimental settings described by the
same equations should exist.

 
\section{Conclusions} 
\label{sec:conclusions} 

We studied the extension of the fluctuation relation
 to open stochastic systems that are not able to equilibrate with
their environments. 

We used the simplest example at hand to test several generalized
fluctuation formulas: a Brownian particle in a confining potential
coupled to non-trivial external baths with different time-scales and
temperatures. Independently of the form of the potential energies, due
to the coupling to the complex environment, the particle is not
able to equilibrate. Its relaxational dynamics is characterized by an
{\it effective temperature}, defined via the modification of the
fluctuation-dissipation relation between spontaneous and induced
fluctuations~\cite{Cukupe}. When no separation of time-scales can be
identified in the bath, the effective temperature is a non-trivial
function of the two times involved. Instead, when the bath evolves in different
time-scales each characterized by a value of a temperature, the
two-time dependent effective temperature is a piece-wise function that
actually takes only these values, each one characterizing the dynamics
of the particle in a regime of times.

Several authors discussed the possibility of introducing the effective
temperature in the fluctuation theorem to extend its domain of
applicability to glassy models driven by external
forces~\cite{Se98,CR04,SCM04,Sasa}. After summarizing our results we shall
discuss how they compare to the proposals and findings in these 
papers.
  
\subsection{Summary of results}

We here examined carefully different
definitions of entropy production rate that are not equivalent when
the {\it effective temperature} is not trivially equal to the ambient
temperature. We found that:

\begin{enumerate}

\item 
The {\sc pdf} of the ``dissipative'' entropy production 
${\cal S}^{diss}_\tau$ that involves the {\it 
frequency dependent} temperature, 
\beq 
\label{concl-seff} {\cal{S}}_\t^{diss}=
\frac{\D\o}{2\p} \sum_{n=-\io}^\io \frac{-i \o_n r_\a(\o_n)
h_\a(\o_n)}{T(\o_n)} = \int_{-\T/2}^{\T/2} dt \int_{-\T/2}^{\T/2} dt'
\, T^{-1}(t-t') \dot r_\a(t) h_\a[\vec r(t')] \ , 
\eeq 
with
$T^{-1}(t)$ the Fourier transform of $1/ T(\o)$, the effective
temperature of the relaxing system, see Eq.~(\ref{eq:Tomega-def}),
verifies the fluctuation theorem exactly for harmonic system.
It also holds for general systems connected to baths with different
 temperaures acting
on widely separated scales.

\item
For nonlinear systems in contact to nonequilibrium baths acting on 
overlapping timescales 
an additional term ${\cal S}^V_\tau$, see Eq.~(\ref{eq:generic-result}),
has to be included in the entropy production to make the fluctuation 
relation hold strictly.
Our numerical simulations suggest that, surprisingly enough,
the effect of this extra `internal' term is even then quite small.

\item 
The {\sc pdf} of ${\cal S}^\Theta_\t$ with 
$\sigma_t^\Theta = W_t/\Th$, $W_t$ the {\it power 
dissipated by the external force} and $\Th$ a free parameter with 
the dimensions of a temperature, {\it does not} satisfy the fluctuation 
theorem in general for any choice of $\Theta$. 
 
The large deviation function, $\z_\Th(p)$, still shows some
interesting features revealing the existence of an effective
temperature.  When the bath has, say, two components acting on
different time scales and with different temperatures, the function
$[\z_\Th(p)-\z_\Th(-p)]/\su_+$ may have different slopes corresponding
to these two temperatures, one at small $p$ and the other at large
$p$. The separation of time-scales of the bath translates into a
separation of scales in the function $[\z_\Th(p)-\z_\Th(-p)]/\su_+$.

When the time scales of the baths are not separated, and one records
the large deviation function for not too large values of $p$ only, the
fluctuation theorem is verified approximately if $\Th$ is suitably
chosen. Note that the value of $\Th$ found in this way is
not equal to the effective temperature $T_{eff}$ that enters the
modified fluctuation--dissipation relation.
Instead, when the time-scales are well separated, 
the two scales in the large deviation 
function are clearly visible and a single fitting parameter is not sufficient
to make the fluctuation theorem hold.

\item
If two time scales are present in the dynamics of a system and the
applied perturbation is periodic with frequency $\O < 1/\t_s$, $\t_s$
being the largest relaxation time, the {\sc pdf} of the power
dissipated over a (large) number of cycles verifies the fluctuation
relation with temperature $T_s=T_{eff}(\O)$. This is probably the
easiest way of detecting the effective temperature by means of the
fluctuation relation.

\end{enumerate} 

These results should apply to driven glassy systems as discussed in
section~\ref{sec:VI} and are indeed consistent with recent numerical
simulations~\cite{ZRA04b}.  Models like the one discussed here have
been recently investigated \cite{Cukuja,AG04,PM04,Po04} to describe
the dynamics of Brownian particles in complex media such as glasses,
granular matter, etc.  Brownian particles are often used as probes in
order to study the properties of the medium (\eg in Dynamic Light
Scattering or Diffusing Wave Spectroscopy experiments).  Moreover,
confining potentials for Brownian particles can be generated using
laser beams \cite{Gr03} and experiments on the fluctuations of the
power dissipated in such systems are currently being performed
\cite{AG04,WSMSE02}.

\subsection{Temperatures}

It is important to summarize the different definitions of effective
temperature we considered and the relations between them. We defined
the effective temperature in the frequency domain in
equation~(\ref{eq:Tomega-def}) as a property of the bath which can also
be measured from the ratio between correlation and response functions
in the frequency domain. As we discussed above, {\it the same}
effective temperature enters the correct definition of entropy
production rate {\it in the frequency domain}, see
equation~(\ref{concl-seff}).  Thus, experiments working in the
frequency domain should observe the same effective temperature from
the fluctuation--dissipation relation and the fluctuation
relation.

In the time domain the situation is slightly more complicated. On the one
hand, the effective temperature obtained from the
fluctuation--dissipation relation {\it in the time domain}, defined
for example by Eq.~(\ref{eq:Tefftime-def}), is {\it not} the
Fourier transform of $T(\o)$. A convolution with the correlation
function is involved in the relation between $T(\o)$ and
$T_{eff}(t)$.  On the other hand, the effective temperature
$T^{-1}(t)$ entering the entropy production is exactly the Fourier
transform of $1/T(\o)$, see again Eq.~(\ref{concl-seff}).
This can give rise to ambiguities when working in the time domain.

Most of these ambiguities disappear as long as the time scales in the
problem are well separated. In this case, on each time scale a well
defined effective temperature can be identified, and this temperature
enters both the fluctuation--dissipation relation and the fluctuation
relation: see {\it e.g.} the curve for $\t=250$ in Fig.~\ref{fig:3b} and the
expression of ${\cal S}_\tau^{diss}$ in the adiabatic approximation,
eq.~(\ref{sigma3adiabatic}).  This is essentially related to the
validity of the adiabatic approximation discussed in
section~\ref{sec:adiabatic}.

The difference is relevant when the time scales of the two baths are
not well separated, and a single effective temperature cannot be
identified, see the curve for $\t=1$ in Fig.~\ref{fig:3b}. In this
case, we found that the fluctuation relation holds with
--approximately-- a single fitting parameter $\Th$ but this
temperature {\it is not clearly related} to the
fluctuation--dissipation temperature {\it in the time domain}. This is
indeed what is observed in numerical simulations on Lennard--Jones
systems~\cite{ZRA04b}.  

Let us remark again that, when applying these results to real glassy
systems in finite dimension, one should take care of the possibility
that the effective temperature has some space fluctuations due to the
heterogeneity of the dynamics~\cite{Cu04}. The extension of our
results to such a situation is left for future work.

\subsection{Discussion}

Several proposals to introduce the effective temperature into
extensions of the fluctuation theorem have recently appeared in the
literature. Let us confront them here to our results.

Sellitto studied the fluctuations of entropy production in a driven
lattice gas with reversible kinetic constraints~\cite{Se98}. 
When coupling this system to an external particle reservoir 
with chemical potential $\mu$, a dynamic crossover from a 
fluid to a glassy phase is found around $\mu_c$. The glassy
nonequilibrium phase is characterized by a violation of the fluctuation
dissipation theorem in which the parametric relation 
between global integrated response and displacement 
yields a line with slope $\mu_{eff}$~\cite{KPS}.
 
One drives this (possibly already out of equilibrium) 
system by coupling two adjacent layers 
of the three dimensional periodic cube
to particle reservoirs at different
chemical potentials, $\mu_+$ and $\mu_-$. 
The former is allowed to assume value corresponding to
the glassy phase, $\mu_+ > \mu_c$, while $\mu_-$ is always 
below $\mu_c$.
The results of 
the Montecarlo simulation are consistent with a 
generalized form of the fluctuation-theorem:
\begin{equation}
\sigma_\t = J_\tau (\mu_{eff}-\mu_-)
\; ,
\label{eq:gen-driven-lattice}
\end{equation} 
where $\sigma_\t$ is the entropy production, $J_\tau$ is the particle
current in the direction of the externally imposed chemical
potential gradient averaged over a time-interval of duration $\tau$;
$\mu_{eff}$ is an effective chemical potential and $\mu_-$ is the
chemical potential of one of the layers.  When the chemical potentials
of the two reservoirs are in the fluid phase, $\mu_{eff}=\mu_+$ and
the usual fluctuation relation holds. Instead, when $\mu_+$ is in the
glassy phase, Sellitto found that Eq.~(\ref{eq:gen-driven-lattice})
holds with $\mu_{eff}$ taking the value appearing in the violation of
fluctuation-dissipation theorem in the aging regime of the {\it
undriven} glassy system at $\mu_+$.

The formula (\ref{eq:gen-driven-lattice}) differs from the ones we 
found to describe the oscillator problem in that in our case,
when translating from temperature to chemical potential, the 
full time-dependent $\mu(t)$ enters. Strictly, we believe that 
this improved definition should also apply to the 
lattice gas model. 
However in the case studied by Sellitto the fast dynamics is an ``intra-cage''
dynamics that likely does not contributes to the current. This is a case
in which the perturbation does not produce dissipation at high frequency
so that the difference arising from $\mu(t)\neq \mu_{eff}$ should be tiny 
in this case (see Sect.~\ref{sec:XI}).

More recently, 
Crisanti and Ritort~\cite{CR04} found that the probability
distribution function of the fluctuations of heat exchanges, $Q$,
between an aging random orthogonal model in its `activated regime' (a
long-time regime in which the energy-density decays as a logarithm of
time) and the heat bath is rather well described by a stationary
Gaussian part and a waiting-time dependent exponential tail towards
small values of $Q$.  Assuming that these events are of two types
(`stimulated' and `spontaneous') they proposed to fit the ratio
between the {\sc pdf} of positive and negative `spontaneous' $Q$'s in the
form of a fluctuation theorem, {\it i.e.} to be proportional to
$e^{-2Q/\lambda}$, and relate $\lambda$ to the effective temperature
of the fluctuation-dissipation relation. They found good agreement.
Crisanti, Ritort and Picco
are currently performing simulations to test this hypothesis in
Lennard-Jones mixtures~\cite{Marco}.

Another development is an attempt to generalize the situation considered
by Crooks. He considered a problem that {\it starts from equilibrium in zero
field} and evolves according to some stochastic dynamic rule in the
presence of an arbitrary applied field~\cite{Crooks} and  found that
the ratio between the probability of a trajectory and its
time-reversed one is given by $e^{-\beta \int_0^{t_{max}} dt h(t)
\dot O(t)}$ with $h(t)$ the time-dependent external field that couples
linearly to the observable $O$. For simplicity, let us focus on
$O=\phi$ with $\phi$ a scalar field characterizing the system. In
\cite{SCM04} the extension of this relation to the initial
non-equilibrium `glassy' case was conjectured. Separating the external
fields $h$ and $\phi$ in their fast and slow
components~\cite{Zannetti}, $h=h_f+h_s$ and $\phi=\phi_f+\phi_s$, one
then proposes that the {\sc pdf}s of the trajectories of the 
slow components satisfy a relation similar to Crooks'
with the temperature replaced by the effective temperature (for a
glassy non-equilibrium system with two correlation
scales~\cite{Cuku93}).
This is indeed very similar to what we have done in this paper.

Finally, let us mention the work of Sasa~\cite{Sasa} where he introduces  
an effective temperature in his definition of entropy production for 
the Kuramoto-Sivashinsky equation.
 
\acknowledgments

F.~Z. would like to thank R.~Di~Leonardo, G.~Gallavotti, and G.~Ruocco 
for many useful discussions and is particularly indebted to 
A.~Giuliani for many technical suggestions and interesting discussion. 
L. F. C. thanks M. Picco for very useful discussions;
she is a member of the Institut Universitaire de 
France and acknowledges financial aid from the European network 
STIPCO.  We all thank ICTP Trieste for hospitality during part of the 
preparation of this work.
 

\appendix 
 
\section{Dirichlet boundary conditions for the white bath} 
\label{app:A} 
 
A second possibility to calculate the functional integral in 
Eq.~(\ref{AAA1}) 
is to impose Dirichlet boundary conditions $z(-\T/2)=z(\T/2)=0$. 
However, in this case it is possible to calculate $\phi(\l)$ only for $m=0$. 
The distribution of $z_t$ is obtained substituting $\r_\o = D(\o) z_\o$ in Eq.~(\ref{AAA3}): 
\beq 
{\cal P}[z_t]= \exp \left[ -\frac{1}{2 \gamma T}  
\int_{-\infty}^\infty \frac{d\omega}{2\pi} \ z_\omega  |D(\o)|^2 \bar z_\omega \right] = 
\exp \left[ -\frac{1}{2 \g T}  
\int_{-\infty}^\infty dt \ z_t   
\left( k^2 + \alpha^2 - 2 i \alpha \g \frac{d}{dt} -\g^2 \frac{d^2}{dt^2}\right) 
\bar z_t \right] \ . 
\eeq 
From Eq.~(\ref{AAA1}) 
\beq 
\langle \exp[- \lambda {\cal S}_\T ] \rangle = 
{\cal N}^{-1} \int dz_t  
\exp \left[ -\frac{1}{2 \g T}  
\int_{-\infty}^\infty dt \ z_t   
\left( k^2 + \alpha^2 - 2 i \alpha \g [1 - 2\lambda \chi_\T(t)]  
\frac{d}{dt} -\g^2 \frac{d^2}{dt^2}\right) 
\bar z_t \right] \ , 
\eeq 
where $\chi_\T(t)$ is the characteristic function of $t\in[-\T/2,\T/2]$. 
At the leading order in $\T$, as the correlation function 
of $z_t$ decays exponentially on a time scale $\t_0=\g k^{-1}$, we can integrate out 
the portion of the trajectory that is outside the interval $[-\T/2,\T/2]$ 
both in the numerator and the denominator, and we simply obtain 
\beq 
\langle \exp[- \lambda {\cal S}_\T ] \rangle = 
 {\cal N}^{-1} \int dz_t  
\exp \left[ -\frac{1}{2 \g T}  
\int_{-\T/2}^{\T/2} dt \ z_t   
\left( k^2 + \alpha^2 - 2 i \alpha \g (1 - 2\lambda)  
\frac{d}{dt} -\g^2 \frac{d^2}{dt^2}\right) 
\bar z_t \right] \ . 
\eeq 
We have then to find the eigenvalues of the operator appearing 
in the integral. This corresponds to find the solution of the equation 
\beq 
J \bar z_t =  
\left(k^2 + \alpha^2 -2i\alpha \g(1 - 2\lambda)\frac{d}{dt}  -  
\g^2 \frac{d^2}{dt^2}\right) \bar z_t = E \bar z_t 
\eeq 
with boundary conditions $\bar z(\T/2)=\bar z(-\T/2)=0$.  
Note that the operator $J$ is Hermitian, thus the eigenvalues are real; 
they are given by the following expression: 
\beq 
E_n(\lambda)=k^2 + 4\alpha^2 \lambda (1-\lambda) +\g^2 \frac{\pi^2 n^2}{\T^2} 
\eeq 
with $n = 0,1,\cdots$. 
For each $n$ the integration is performed on one complex variable and 
we get 
\beq 
\langle \exp[- \lambda {\cal S}_\T ] \rangle = 
 {\cal N}^{-1} \int dz_t  
\exp \left[ -\frac{1}{2 \g T}  
\int_{-\T/2}^{\T/2} dt \ z_t   
J \bar z_t \right] = \prod_{n=0}^\infty \frac{E_n(0)}{E_n(\lambda)} 
\eeq 
recalling that the constant ${\cal N}$ is simply the numerator calculated 
in $\lambda=0$. Finally we obtain, defining $\o=n\pi/\T$, 
\beq 
\label{phi1bagno} 
\phi(\lambda)= \lim_{\T \rightarrow \infty} \frac{1}{\T}
\sum_{n=0}^{\infty} \ln \frac{E_n(\lambda)}{E_n(0)} = \int_0^\infty
\frac{d\o}{\pi} \ln \left[ 1 + \frac{ 4\alpha^2 \lambda (1-\lambda) }{
k^2 + \g^2 \o^2 } \right] \eeq The latter expression verifies
obviously the fluctuation theorem. Moreover, in the $m=0$ case
Eq.~(\ref{phi1bagnoFourier}) is equal to Eq.~(\ref{phi1bagno}), as one
can check using suitable changes of variable in the integral.
In this simple
case, $\zeta(p)$ can be computed exactly. Starting from
Eq.~(\ref{phi1bagno}) one has 
\beq \phi'(\lambda)=\int_0^\infty
\frac{d\o}{\pi} \frac{4 \alpha^2 (1-2\lambda)} {\g^2 \o^2 + k^2+ 4
\alpha^2 \lambda(1-\lambda)} = \frac{2\alpha^2 (1-2\lambda)} {\g
\sqrt{k^2+ 4 \alpha^2 \lambda(1-\lambda)}} \ , \eeq and, recalling
that $\phi(0)=0$, \beq \phi(\lambda)=\int_0^\lambda d\mu \ \phi'(\mu)
= \g^{-1} \big[ \sqrt{k^2+ 4 \alpha^2 \lambda(1-\lambda)} - k \big] \
.  \eeq 
The function $\zeta(p)$ is defined by \beq \zeta(p) =
\min_\lambda [\lambda p \s_+ - \phi(\lambda)] = \lambda^* p \s_+ -
\phi(\lambda^*) \ , \eeq where $\s_+=2\a^2/(\g k)$ and $\lambda^*$ is
defined by $\phi'(\lambda^*)= p \s_+$; hence, 
\beq p= \frac{ k
(1-2\lambda^*)}{\sqrt{k^2+ 4 \alpha^2 \lambda^* (1-\lambda^*)}}
\hskip10pt \Rightarrow \hskip10pt \lambda^* = \frac{1}{2} \left[ 1 - p
\sqrt{\frac{\alpha^2 + k^2} {\alpha^2 p^2 + k^2}} \right] \ , 
\eeq 
and finally 
\beq \zeta(p)=\g^{-1} \left\{ k + \frac{\alpha^2 p}{k} \left[
1 - p \sqrt{\frac{\alpha^2 + k^2} {\alpha^2 p^2 + k^2}} \right] - k
\sqrt{\frac{\alpha^2 + k^2} {\alpha^2 p^2 + k^2}} \right\} \ .  
\eeq
From the latter expression it is easy to verify that 
\beq
\zeta(p)-\zeta(-p) = \frac{2\alpha^2 p}{k} = p \s_+ \ , 
\eeq 
as stated
by the FT.  Defining $\t_0 = \g/k$, the relaxation time of the
correlation function of $z_t$, and $\s_0 = \s_+ \t_0/2 = \a^2/k^2$,
the (adimensional) entropy production over a time $\t_0/2$, we obtain
\beq 
\z(p)=\t_0^{-1} \left[ 1 + p \s_0 - \sqrt{(1+\s_0)(1+p^2 \s_0)}
\right] \ .  
\eeq
 
%

\section{Fluctuation theorem for 
many equilibrium baths at different temperatures}
\label{app:III} 
 
We compute the function $\phi(\l)$ in the case in which the driven
oscillator is coupled to $N$ colored baths with generic memory functions and
in equilibrium at different temperatures. The violation of the
fluctuation-dissipation theorem for the relaxing particle in such an
environment was discussed in Sect.~\ref{subsec:noneq-baths}.
As discussed there, the equations are mathematically equivalent to the ones
discussed in Sect.~\ref{sec:II}; thus the strategy as 
well as many details of the calculation are the same as in 
this Section.

The equations of motion are  
\beq 
\label{motionNbaths} 
m \ddot z_t +  
\sum_{i=1}^N \Iint ds \ g_i(t-s) \dot z_s  
= -\kappa z_t +  
\sum_{i=1}^N \rho_{it} \ , 
\eeq 
with $\kappa = k - i\alpha$. The thermal noises satisfy 
\beq 
\label{bathsdef} 
\begin{split} 
&\langle \rho_{it} \rho_{j0} \rangle =  
\langle \bar\rho_{it} \bar\rho_{j0} \rangle = 0 \ ,\\ 
&\langle \rho_{it} \bar\rho_{j0} \rangle =  
\delta_{ij} T_i \nu_i(t) \ . 
\end{split} 
\eeq  
By causality, the functions $g_i(t)$ must vanish for $t <0$.  As 
the baths are in equilibrium at temperature $T_i$, the 
functions $\nu_i(t)$ and $g_i(t)$ are related by 
Eq.~(\ref{FDTbathTD}):  
\beq 
\label{f&g} 
\begin{split} 
& \nu_i(t) = T_i [ g_i(t) + g_i(-t) ] = T_i g_i(|t|) \ , \\ 
& T_i g_i(t) = \theta(t) \nu_i(t) \ . 
\end{split} 
\eeq 
In the frequency domain Eq.~(\ref{motionNbaths}) becomes 
\beq 
z_\omega = \frac{\sum_i \rho_{i\omega}} 
{-m \omega^2 + \kappa - i\omega \sum_i g_i(\omega)} 
\equiv \frac{\sum_i \rho_{i\omega}}{D(\omega)} \ , 
\eeq 
where $D(\omega) = -m \omega^2 + \kappa - i\omega \sum_i g_i(\omega)$. 
 
The dissipated power is given by  
\beq  
\frac{dH}{dt} = 2\alpha \; \im \dot z_t \bar z_t -2\re 
\sum_i \int_{-\infty}^\infty ds \ g_i(t-s) \dot z_t \dot{\bar z}_s + 2 
\re \sum_i \dot z_t \bar \rho_{it} = W_t - \sum_i \widetilde W_{it} 
\;  ,  
\eeq  
where as in the previous cases $W_t = 2\alpha \im \dot z_t \bar 
z_t $ is the power injected by the external force and $\widetilde 
W_{it} = 2\re \int_{-\infty}^\infty ds \ g_i(t-s) \dot z_t \dot{\bar 
z}_s - 2 \re \dot z_t \bar \rho_{it}$ is the power extracted by the 
$i$-th bath. 

The first definition of entropy production rate, Eq.~(\ref{sigma1}),
gives (in the following, 
we will always substitute $\frac{\D\o}{2\pi} \sum_{n=-\io}^\io 
\rightarrow \int_{-\io}^\io \frac{d\o}{2\pi}$ as the error is $O(1)$ 
for $\T \rightarrow \io$, see section~\ref{sec:II}): 
\beq 
{\cal S}^\Th_\T = -\int_{-\infty}^\infty \frac{d\omega}{2\pi}  
\frac{2\alpha\omega|z_\omega|^2}{\Th} \ . 
\eeq 
Substituting $z_\o = \sum_i \r_{i\o} / D(\o)$, we obtain 
\beq 
\langle \exp[- \lambda {\cal S}^\Th_\T ] \rangle = {\cal N}^{-1} 
\int d\rho_{i\omega} \  
\exp \left[ - \int_{-\infty}^\infty \frac{d\omega}{2 \pi} 
 \sum_{ij} \rho_{i\omega} A_{ij}^\lambda(\omega) \bar \rho_{j\omega} 
\right] \ , 
\eeq 
where $A^\lambda(\omega)$ is a $N\times N$ {\it real} matrix which elements are 
given by 
\beq 
A_{ij}^\lambda(\omega) = 
\frac{\delta_{ij}}{T_i \nu_i(\omega)} - \frac\lambda{|D(\omega)|^2} 
\frac{2\alpha\omega}{\Th} \ . 
\eeq 
Then, 
\beq 
\phi_\Th(\lambda)=\lim_{\T\rightarrow \infty} \T^{-1} \ln 
\prod_{n=-\infty}^\infty \frac{\det A^\lambda(\omega_n)} 
{\det A^0(\omega_n)} 
= \int_{-\infty}^\infty \frac{d\omega}{2 \pi} \  
\ln \left[ 
\frac{\det A^\lambda(\omega)}{\det A^0(\omega)} \right] \ . 
\eeq 
The determinant of a matrix of the form $A^\l_{ij}=c_i^{-1} \d_{ij} + \l b$ satisfies
the relation
\beq
\frac{\det A^\lambda}{\det A^0} = 1 + \l b \sum_i c_i \; ;
\eeq
we finally obtain 
\beq
\label{phi1Nbaths}
\phi_\Th(\lambda)= 
 \int_{-\infty}^\infty \frac{d\omega}{2 \pi} \  
\ln \left[ 1 - \frac{2 \alpha \lambda \omega \sum_i \frac{T_i}{\Th} \nu_i(\omega)}  
{|D(\omega)|^2}\right] \ . 
\eeq 
In general, {\it it does not exist a choice of $\Th$ such that $\phi_\Th(\l)$ 
verifies the fluctuation theorem}, \ie 
$\phi_\Th(\l) \neq \phi_\Th(1-\l)$. 
 
For the second definition, given by Eq.~(\ref{sigma3}), the 
computation is identical to the one of the previous section with the 
substitution $\Th \rightarrow T(\o)$, where $T(\omega)$ is 
given by Eq.~(\ref{TeffNbaths}). The result is then  
\beq 
\phi_{diss}(\lambda) =  
\int_{-\infty}^\infty \frac{d\omega}{2 \pi} \ \ln 
\left[ 1 - \frac{2 \alpha \lambda \omega \sum_i T_i \nu_i(\omega)} 
{T_{eff}(\omega) |D(\omega)|^2}\right]= \int_{-\infty}^\infty 
\frac{d\omega}{2 \pi} \ \ln \left[ 1 - \frac{2 \alpha \lambda \omega 
\sum_i \nu_i(\omega)} {|D(\omega)|^2} \right] \ .
\eeq  
Observing that 
\beq 
\label{Dsymm} 
\begin{split} 
&D(-\omega)=\overline{D(\omega)} - 2 i\alpha \ , \\ 
&|D(-\omega)|^2 = |D(\omega)|^2 - 2 \alpha \omega \sum_i \nu_i(\omega) \ ,
\end{split} 
\eeq 
and  it is now easy 
to show that $\phi_{diss}(\lambda)=\phi_{diss}(1-\lambda)$.

 
\section{Entropy production of the thermal baths}
\label{app:B} 

We will discuss here a different definition of entropy production rate based on the
power {\it extracted by the thermal bath} instead of the one injected by the driving
force. The two differ by a total derivative if there is only one bath, so their asymptotic
distributions should be identical if boundary terms can be neglected.
However Van Zon and Cohen \cite{VzC04} showed in a particular case that this argument 
is not correct, see also \cite{BGGZ05}. 

If there are many baths equilibrated at different temperature, 
the study of the entropy production extracted by each bath
allows to separate the different contributions to the total entropy production
weighting each one with the right temperature.
We will first discuss the case of a single bath, and later we perform the computation
for the general case. 
We will show that the entropy 
production rate defined in this way satisfies the fluctuation theorem as outlined in~\cite{Ga04};
unfortunately, in our computation we neglect all boundary terms so we cannot 
check if these terms modify the asymptotic distribution, as observed in \cite{VzC04,BGGZ05}.
Anyway, the modification can be proven to be eventually relevant only for 
$|p| > 1$~\cite{VzC04,BGGZ05}, 
so the results we will discuss should hold {\it at least} for $|p|\leq 1$.

\subsection{One bath}
 
The entropy production rate of the bath is defined as: 
\beq 
\label{sigmatilde1bagno} 
\s^{bath}_t = \beta \widetilde W_t = \beta  
[ 2 \gamma \dot z_t \dot{\bar z}_t - 2 \re \dot z_t \bar \r_t ] =  
\sigma_t + \beta\frac{dH}{dt}  
\ , 
\eeq 
where $\widetilde W_t$ has been defined in section~\ref{sec:II} and 
is the power extracted from the system by the thermostat. 
Since $\sigma_t$ and $\widetilde \sigma_t$ differ only by a 
total derivative [see Eq.~(\ref{sigmatilde1bagno})], for large 
$\T$ we have 
\beq 
\label{DDD1} 
{\cal S}^{bath}_\T= {\cal S}_\T +\beta [ H(\T/2) - H(-\T/2) ] 
= {\cal S}_\T + \beta \D H \ . 
\eeq 
The first term in the r.h.s. is $O(\T)$ and has fluctuations 
$O(\sqrt{\T})$, while the second term has zero average and its 
fluctuations are also $O(1)$. 
On this ground one usually neglects the second 
term and concludes that  
\beq 
\z_{bath}(p)=\z(p) 
\; ;
\eeq 
the two definitions of entropy production rate are equivalent and both 
distributions verify the fluctuation theorem.
Van Zon and Cohen studied a model 
very similar to the one considered here, but where the distribution of 
${\cal S}_\T$ is Gaussian while the distribution of $\D H$ shows 
exponential tails. In such a case, the large fluctuations of 
${\cal S}^{bath}_\T$ are dominated by the distribution of $\D H$ 
even if $\langle ({\cal S}_\T - \T \s_+)^2 \rangle \gg \langle ( \D H )^2 
\rangle$, see Ref.~\cite{VzC04,BGGZ05} and references therein. The modification
of $\z_{bath}(p)$ occurs for $|p|>1$; then, for $|p|\leq 1$, $\z_{bath}(p)$
still verifies the fluctuation relation.

\subsection{Many baths}
\label{app:B1}

If the system is coupled to many equilibrated baths,
in addition to the definitions of entropy production rate
given by Eqs.~(\ref{sigma1}) and (\ref{sigma3}),
a generalization of Eq.~(\ref{DDD1}) is possible, 
given by the sum of the power dissipated by each bath divided by the
corresponding temperature:  
\beq
\label{sigma2} 
\sigma^{baths}_t = \sum_{i=1}^N \frac{\widetilde W_{it}}{T_i} 
\; . 
\eeq  
This quantity takes into account heat exchanges between the 
baths, and its average value does not vanish at $\a = 0$, as we shall 
see in the following. 
 
Let us compute $\phi_{baths}(\l)$. We have from Eq.~(\ref{sigma2}): 
\beq 
\begin{split} 
{\cal S}^{baths}_\T =  
\int_{-\T/2}^{\T/2} dt \ \s^{baths}_t =  
&\int_{-\infty}^\infty \frac{d\omega}{2\pi} \re \sum_i  
\frac2{T_i} \Big[ 
\omega^2 |z_\omega|^2 g_i(\omega) + i \omega z_\omega \bar \rho_{i\omega}  
\Big]\\ 
=&\int_{-\infty}^\infty \frac{d\omega}{2\pi}  
\left[ \frac{\omega^2 \big|\sum_i \rho_{i\omega} \big|^2  
\sum_j \frac{ \nu_j(\omega)}{T_j} }{|D(\omega)|^2}  
+ \sum_{ij} \rho_{i\omega} \bar\rho_{j\omega}  
\left( \frac{i\omega}{D(\omega) T_j} - \frac{i\omega}{\overline{D(\omega)}T_i} 
\right)\right] \ . 
\end{split} 
\eeq 
If we define the functions  
\beq 
\begin{split} 
&p(\omega)=i \omega  \overline{D (\omega)} \ , \\ 
&F(\omega)=\omega^2 \sum_i\frac{\nu_i(\omega)}{T_i} \ , 
\end{split} 
\eeq 
we obtain 
\beq 
\langle \exp[- \lambda {\cal S}^{baths}_\T ] \rangle = {\cal N}^{-1} 
\int d\rho_{i\omega} \  
\exp \left[ - \int_{-\infty}^\infty \frac{d\omega}{2 \pi} 
 \sum_{ij} \rho_{i\omega} A_{ij}^\lambda(\omega) \bar \rho_{j\omega} 
\right] \ , 
\eeq 
where $A^\lambda(\omega)$ is a $N\times N$ matrix which elements are 
given by 
\beq 
A_{ij}^\lambda(\omega) = \overline{A_{ji}^\lambda(\omega)} = 
\frac{\delta_{ij}}{T_i \nu_i(\omega)} + 
\frac\lambda{|D(\omega)|^2}\left[ F(\omega) + 
\frac{p(\omega)}{T_j}+\frac{\overline{p(\omega)}} 
{T_i} \right] \ . 
\eeq 
Then, 
\beq 
\phi_{baths}(\lambda)=\lim_{\T\rightarrow \infty} \T^{-1} \ln 
\prod_{n=-\infty}^\infty \frac{\det A^\lambda(\omega_n)} 
{\det A^0(\omega_n)} 
= \int_{-\infty}^\infty \frac{d\omega}{2 \pi} \  
\ln \left[ 
\frac{\det A^\lambda(\omega)}{\det A^0(\omega)} \right] \ . 
\eeq 
The matrix $A$ has the following form: 
\beq 
A \sim \left( 
\begin{matrix} 
c_i^{-1} + \mu b_{ii} & \cdots & \mu b_{ij} \\ 
\vdots & \ddots & \vdots \\ 
\mu b_{ji} & \cdots & c_j^{-1} + \mu b_{jj} 
\end{matrix} 
\right) \ , 
\eeq 
where $\mu=\lambda/|D(\omega)|^2$, $c_i=T_i \nu_i(\omega)$ and  
$b_{ij}= F(\omega) + \frac{p(\omega)}{T_j}+\frac{\overline{p(\omega)}}{T_i}$. 
Its determinant is an order $N$ polynomial in $\mu$ of the following form: 
\beq 
\label{detexp} 
\frac{\det A^\lambda}{\det A^0} =  
1 + \mu \sum_i  c_i b_{ii} 
+ \mu^2 \sum_{i<j} c_i c_j \left| 
\begin{matrix} 
b_{ii} & b_{ij} \\ 
b_{ji} & b_{jj} 
\end{matrix} \right| 
+ \mu^3 \sum_{i<j<k} c_i c_j c_k \left| 
\begin{matrix} 
b_{ii} & b_{ij} & b_{ik} \\ 
b_{ji} & b_{jj} & b_{jk} \\ 
b_{ki} & b_{kj} & b_{kk} 
\end{matrix} \right| + \cdots \ . 
\eeq 
Let us compute the coefficients explicitly. 
We will define $T_{ij}$ by $T_{ij}^{-1}=T_i^{-1}-T_j^{-1}$. 
The coefficient of $\lambda^2$ is given by a sum of determinants of the form 
\beq 
\left| 
\begin{matrix} 
F + \frac{p}{T_i}+\frac{\overline{p}}{T_i} &  
F + \frac{p}{T_i}+\frac{\overline{p}}{T_j} \\ 
F + \frac{p}{T_j}+\frac{\overline{p}}{T_i} &  
F + \frac{p}{T_j}+\frac{\overline{p}}{T_j} 
\end{matrix} \right|= 
\left| 
\begin{matrix} 
F + \frac{p}{T_i}+\frac{\overline{p}}{T_j} &  
\frac{\overline{p}}{T_{ji}} \\ 
F + \frac{p}{T_j}+\frac{\overline{p}}{T_i} &  
\frac{\overline{p}}{T_{ji}} 
\end{matrix} \right|= 
\left| 
\begin{matrix} 
\frac{p}{T_{ij}} &  
0 \\ 
F + \frac{p}{T_j}+\frac{\overline{p}}{T_i} &  
\frac{\overline{p}}{T_{ji}} 
\end{matrix} \right|= 
-\frac{|p|^2}{\big( T_{ij} \big)^2} \ , 
\eeq 
where we first subtracted the first column to the second column,  
and then subtracted the second row to the first row. 
We want now to show that all the coefficients of the higher powers of $\lambda$ 
vanish. Consider for example the coefficient of $\lambda^3$. It has the form 
\beq 
\left| 
\begin{matrix} 
F + \frac{p}{T_i}+\frac{\overline{p}}{T_i} &  
F + \frac{p}{T_i}+\frac{\overline{p}}{T_j} & 
F + \frac{p}{T_i}+\frac{\overline{p}}{T_k} \\ 
F + \frac{p}{T_j}+\frac{\overline{p}}{T_i} &  
F + \frac{p}{T_j}+\frac{\overline{p}}{T_j} & 
F + \frac{p}{T_j}+\frac{\overline{p}}{T_k} \\ 
F + \frac{p}{T_k}+\frac{\overline{p}}{T_i} &  
F + \frac{p}{T_k}+\frac{\overline{p}}{T_j} & 
F + \frac{p}{T_k}+\frac{\overline{p}}{T_k} 
\end{matrix} \right|= 
\left| 
\begin{matrix} 
F + \frac{p}{T_i}+\frac{\overline{p}}{T_i} &  
\frac{\overline{p}}{T_{ji}} & 
\frac{\overline{p}}{T_{ki}} \\ 
F + \frac{p}{T_j}+\frac{\overline{p}}{T_i} &  
\frac{\overline{p}}{T_{ji}} & 
\frac{\overline{p}}{T_{ki}} \\ 
F + \frac{p}{T_k}+\frac{\overline{p}}{T_i} &  
\frac{\overline{p}}{T_{ji}} & 
\frac{\overline{p}}{T_{ki}} 
\end{matrix} \right|= 
\left| 
\begin{matrix} 
\frac{p}{T_{ik}} &  
0 & 
0 \\ 
\frac{p}{T_{jk}} &  
0 & 
0 \\ 
F + \frac{p}{T_k}+\frac{\overline{p}}{T_i} &  
\frac{\overline{p}}{T_{ji}} & 
\frac{\overline{p}}{T_{ki}} 
\end{matrix} \right|=0 \ , 
\eeq 
where we subtracted the first column to the second and third column, and then subtracted 
the third row to the first and second row. The same argument applies to all the other 
coefficients up to order $N$. Finally, we get 
\beq 
\begin{split} 
\frac{\det A^\lambda(\omega)}{\det A^0(\omega)} &=  
1 + \frac\lambda{|D(\omega)|^2} \sum_i  T_i \nu_i(\omega)  
\left[ F(\omega) + \frac{p(\omega)}{T_i}+\frac{\overline{p(\omega)}}{T_i} \right] 
- \frac{\lambda^2 |p(\omega)|^2}{|D(\omega)|^4}  
\sum_{i<j} \frac{ T_i T_j \nu_i(\omega) \nu_j(\omega)}{\big( T_{ij} \big)^2} \\ 
&= 1-\frac{2\alpha\omega\lambda \sum_i \nu_i(\omega)}{|D(\omega)|^2} 
 + \frac{\lambda (1-\lambda)}{|D(\omega)|^2}  
\sum_{i<j} \frac{ T_i T_j \nu_i(\omega) \nu_j(\omega)}{\big( T_{ij} \big)^2} \ , 
\end{split} 
\eeq 
and 
\beq 
\phi_{baths}(\l)=\int_{-\io}^{\io} \frac{d\o}{2\p} \ln \left[ 1-\frac{2\alpha\omega\lambda  
\sum_i \nu_i(\omega)}{|D(\omega)|^2} 
 + \frac{\lambda (1-\lambda)}{|D(\omega)|^2}  
\sum_{i<j} \frac{ T_i T_j \nu_i(\omega) \nu_j(\omega)}{\big( T_{ij} \big)^2} \right] \ . 
\eeq 
The first term in the logarithm is proportional to $\a$ and is related to the power injected
by the external force, while the second term accounts for heat exchanges between the baths and
does not vanish at $\a=0$.
Finally, observing that 
\beq 
\begin{split} 
&D(-\omega)=\overline{D(\omega)} - 2 i\alpha \ , \\ 
&|D(-\omega)|^2 = |D(\omega)|^2 - 2 \alpha \omega \sum_i \nu_i(\omega) \ ,
\end{split} 
\eeq 
and using the same trick we already used above it is 
easy to show that $\phi_{baths}(\lambda)= \phi_{baths}(1-\lambda)$. 
Thus $\z_{baths}(p)$ should verify the fluctuation relation {\it at least} for $|p|\leq 1$,
if the contribution of boundary terms is not negligible.
This result is of interest for the study of heat conduction and is similar to the one
discussed in Ref.~\cite{Ga04}.


\section{Correlation functions of the harmonic oscillator coupled to two baths} 
\label{app:C} 
 
In the harmonic case, ${\cal V}(|z|^2)=\frac{k}{2} |z|^2$, the correlation function 
of a variable $z_t$, whose time evolution is given by Eq.~(\ref{motion2baths}),
can be computed analytically \cite{Cukuja}. 
In the frequency domain, Eq.~(\ref{motion2baths}) reads: 
\beq 
z_\omega = \frac{\r^f_{\o} + \r^s_{\o}} 
{ \kappa - i\omega \gamma_f - \frac{i\omega \gamma_s} 
{1-i\omega\tau_s}} \equiv \frac{\r^f_{\o} + \r^s_{\o}}{D(\omega)} \ , 
\eeq 
where $D(\omega) = \kappa - i\omega \gamma_f -  
\frac{i\omega \gamma_s}{1-i\omega\tau_s}$. 
Recalling that 
$\langle \r^f_{\o} \r^f_{\o'} \rangle =  
4\pi \gamma_f T_f \delta(\omega +\omega')$  
and $\langle \r^s_{\o} \r^s_{\o'} \rangle = \frac{4\pi \gamma_s T_s} 
{1+\omega^2\tau_s^2} \delta(\omega + \omega')$, and defining $C(\omega)$ from 
$\langle z_\omega z_\omega' \rangle = 2\pi \delta(\omega + \omega') C(\omega)$ 
we get 
\beq 
\begin{split} 
&R(\omega)=\frac{dz_\omega}{d\r^f_{\omega}}=\frac{1} 
{D(\omega)} \ , \\ 
&C(\omega) = \frac{ 2 \gamma_f T_f + \frac{2 \gamma_s T_s}{1+\omega^2\tau_s^2}} 
{\left| D(\omega) \right|^2} \ . 
\end{split} 
\eeq 
The function 
$(1-\omega \tau_s) D(\omega)$ is a polynomial in $\omega$ and its zeros 
are given by $\omega=- i \gamma_\pm$ where 
\beq 
\gamma_\pm = \frac{1}{2 \gamma_f \tau_s} \left[ (\kappa \tau_s + \gamma_f 
+\gamma_s ) \pm \sqrt{ (\kappa \tau_s + \gamma_f +\gamma_s )^2  
- 4 \kappa \tau_s \gamma_f } \right] \ , 
\eeq 
and $\re \gamma_\pm > 0$. 
The response function is then given by 
\beq 
R(t)=\int_{-\infty}^\infty \frac{d\omega}{2 \pi} e^{-i\omega t}  
\frac{1}{D(\omega)} = \frac{\theta(t)}{\gamma_f \tau_s} 
\left[ \frac{1 - \gamma_+ \tau_s}{\gamma_--\gamma_+} e^{-\gamma_+ t} 
+\frac{1 - \gamma_- \tau_s}{\gamma_+-\gamma_-} e^{-\gamma_- t}\right] \ , 
\eeq 
and the correlation function is given by 
\beq 
C(t)=\frac{1}{(\gamma_f \tau_s)^2} \left[ 
\frac{\gamma_f T_f (1-\gamma_+^2 \tau_s^2) + \gamma_s T_s} 
{( \gamma_- -\gamma_+ ) (\bar\gamma_-+\gamma_+)\re\gamma_+ } 
e^{-\gamma_+ t} +  
\frac{\gamma_f T_f (1-\gamma_-^2 \tau_s^2) + \gamma_s T_s} 
{( \gamma_+ - \gamma_- )(\bar\gamma_+ + \gamma_-) \re\gamma_- }  
e^{-\gamma_- t} \right] \ . 
\eeq 
In the case $\alpha = 0$, and in the limit $\gamma_f \ll \gamma_s \ll k \tau_s$  
where the time scales of the two baths are well separated, one obtains 
\beq 
\begin{split} 
& \gamma_+ \sim \frac{k \tau_s + \gamma_s}{\gamma_f \tau_s} \ , \\ 
& \gamma_- \sim \frac{1}{\tau_s} \left( 1 - \frac{\gamma_s}{\gamma_s + k\tau_s} \ , 
\right) 
\end{split} 
\eeq 
and 
\beq 
\begin{split} 
& C(t) = \frac{T_s \gamma_s \tau_s}{(k\tau_s + \gamma_s)^2} e^{-t/\tau_s}+ 
\frac{T_f \tau_s}{k\tau_s + \gamma_f} 
e^{- \frac{k \tau_s + \gamma_s}{\gamma_f \tau_s} t} \ , \\ 
& R(t) = \theta(t) \left[ \frac{\gamma_s}{(k\tau_s + \gamma_s)^2} e^{-t/\tau_s}+ 
\frac{1}{\gamma_f} 
e^{- \frac{k \tau_s + \gamma_s}{\gamma_f \tau_s} t} \right] \ . 
\end{split} 
\eeq 
From the latter expressions it is easy to check that one has 
$R(t) \sim - \beta_f \theta(t) \dot C(t)$ for short times ($t\ll\tau_s$) 
and $R(t) \sim - \beta_s \theta(t) \dot C(t)$ for large times ($t\sim\tau_s$). 
The same behavior is found in the limit of small dissipation (small 
$\alpha$), as one can check  
plotting the exact expression for the functions $R(t)$ and $C(t)$. 

 
\section{The expression of $\sigma^{diss}$ in the adiabatic approximation} 
\label{app:D} 
 
We start from the expression~(\ref{sigma3OK}) for $\sigma^{diss}_t$ and from Eq.~(\ref{Tstar}).  
Then (remember that $\int_{-\io}^t ds \ \delta(t-s) = \theta(0) = 1/2$  
in our convention): 
\beq 
\int_{-\io}^t dt' \ T^{-1}(t-t') \dot z_t \bar z_{t'} =  
\frac{\dot z_t \bar z_t}{2T_f} + \frac{\g_s}{2 \Omega T_f \g_f (\t_s)^2}  
\left(1-\frac{T_s}{T_f}\right) \int_{-\io}^t dt' \ e^{-\Omega (t-t')} \dot z_t \bar z_{t'} \ . 
\eeq 
We substitute $z_t = H_t + w_t$ and neglect all the terms proportional to $H_t w_t$: 
indeed, such terms vanish when $\sigma^{diss}_t$ is integrated over time intervals of the order 
of $\t_s$, as, on such time scales, $\langle w_t \rangle = 0$ while $H$ is constant. 
The first term gives then 
\beq 
\frac{\dot z_t \bar z_t}{2T_f} = \frac{\dot H_t \bar H_t + \dot w_t \bar w_t}{2T_f} \ . 
\eeq 
In the second term, as $\Omega \sim 1/\t_s$, we approximate  
$\int_{-\io}^t dt' \ e^{-\Omega (t-t')} \bar z_{t'} \sim H_t/\Omega$ to obtain 
\beq 
\frac{\g_s}{2 \Omega^2 T_f \g_f (\t_s)^2}  
\left(1-\frac{T_s}{T_f}\right) \dot H_t \bar H_t \sim  
\frac{1}{2T_s} \left(1-\frac{T_s}{T_f}\right) \dot H_t \bar H_t \ . 
\eeq 
The (imaginary part of the) sum of these two terms times $4 \a$ gives  
Eq.~(\ref{sigma3adiabatic}). 



\begin{thebibliography}{99}
\bibitem{comment1} We shall not address in this paper the problem of the
identification of the phase space contraction rate with the entropy production
rate in dynamical systems.
\bibitem{comment2}
The discussion in these paragraphs has to be modified
in the $k=0$ limit in which the particle 
does not have a confining potential and diffuses.
\bibitem{newcomment}
This
is also observed in the simulation, because the average of ${\cal S}^V_\t$
vanishes and the variance of ${\cal S}^V_\t$ does not grow with $\t$.
\end{thebibliography}
\end{document}